# The Role of Interfacial Morphology in $Cu_2O/TiO_2$ and Band Bending: Insights from Density Functional Theory

Mona Asadinamin,[1] Aleksandar Živkovic[2,3], Nora H. De Leeuw[2], and Steven P. Lewis[1]

[1] Department of Physics and Astronomy, University of Georgia, Athens, Georgia 30602, US

[2] Department of Earth Sciences, Utrecht University, Princetonlaan 8a, 3548CB Utrecht, The Netherlands

[3] Institut für Theoretische Physik und Astrophysik, Christian-Albrechts-Universität zu Kiel, Leibnizstraße 15, 24118 Kiel, Germany

Corresponding author: a.zivkovic@uu.nl, splewis@uga.edu

**Abstract**

Photocatalysis, a promising solution for environmental challenges, relies on the generation and utilization of photogenerated charge carriers within photocatalysts. However, recombination of these carriers often limits efficiency. Heterostructures, especially $Cu_2O/TiO_2$, have emerged as effective solutions to enhance charge separation. This study systematically explores the effect of interfacial morphologies on the band bending within $Cu_2O/TiO_2$ anatase heterostructures, employing density functional theory (DFT). Through this study, eight distinct interfaces are identified and analyzed, revealing a consistent staggered-type band alignment. Despite variations in band edge positions, a systematic charge transfer from $Cu_2O$ to $TiO_2$ is observed across all interfaces. The proposed band bending configurations would suggest enhanced charge separation and photocatalytic activity under ultraviolet illumination due to a Z-scheme configuration. This theoretical investigation provides valuable insights into the interplay between interfacial morphology, band bending, and charge transfer, for advancing the understanding of fundamental electronic mechanisms in heterostructures.

**Introduction**

Photocatalysis, a promising solution to environmental challenges like air purification and wastewater treatment, operates via the absorption of photons to create charge carriers (electron-hole pairs) within a photocatalyst. These photogenerated charge carriers are then anticipated to participate in surface redox reactions that fuel the photocatalytic process. A challenge, however, is the tendency of the photogenerated electrons and holes to recombine which deactivates the charge carriers before they can contribute to the desired photocatalytic reactions. Particularly in single photocatalysts, the lack of a suitable mechanism to efficiently separate and transport these charge carriers often leads to high recombination rates. Various strategies have been proposed to overcome these limitations, such as surface modification[1-3], metal or non-metal doping[4, 5], etc. Among these, heterostructures have emerged as effective solutions by significantly enhancing efficiency via charge separation[6-9]. Key benefits include improved separation of photogenerated charge carriers, achieved through staggered band alignments across different materials, leading to reduced recombination rates and longer charge carrier lifetimes. Furthermore, materials like $TiO_2$, while exhibiting excellent photocatalytic performance, suffer from a wide band gap, limiting their effectiveness under visible light. By forming heterostructures, it is possible to combine such compounds with narrow band gap materials to enhance the utilization of the visible light spectrum.

In a heterostructure, when two semiconductor materials meet at the junction, their differing energy band gaps and band edge positions cause an initial discontinuity in the fermi energy. To reach equilibrium, electrons and holes migrate in opposing directions. This charge transport gives rise to an interfacial space charge region, stemming from the electric field, which consequently results in band bending[10]. Depending on the relative band edges and configuration of band bending, semiconductor interfaces can be organized into four types of heterojunctions: straddling gap (type I), staggered gap (type II), Z-scheme, or broken gap (type III)[11]. Type II and Z-scheme interfaces can reduce electron-hole recombination and increase the migration of specific charge carriers to the semiconductor surface, enhancing photocatalytic reactions. Identifying the band bending configuration is key to predict the fundamental photocatalytic mechanisms of heterostructures.

In evaluating the band bending at interfaces, the morphology of materials can influence their interfacial band structure and band bending properties[12-15]. With advancements in novel growth techniques, such as molecular-beam epitaxy, epitaxial interfaces of exceptional quality can be fabricated[16, 17]. This is achievable not just between lattice-matched semiconductors but also between materials with considerable differences in their lattice constants[18, 19]. When these lattice mismatches are present, uniform lattice strain can accommodate them if the layers are sufficiently thin[20]. These strains introduce alterations to electronic properties, offering enhanced versatility in designing semiconductor devices[20, 21]. The interplay between structural and electronic properties at interfaces can be explored both experimentally and theoretically, with each technique coming with their own set of opportunities and obstacles. Density functional theory (DFT) calculations as well as experimental methods, like in situ X-ray photoelectron spectroscopy and scanning transmission electron microscopy (TEM), reveal that structural changes, including planar defects[22], elastic strain[23], tensile strain[24], and introduction of impurities and vacancies[25] affect the electronic transport properties of the interface. Specifically, these changes influence band edge offsets, enabling tunable band bending properties for photocatalytic applications. However, from a computational point of view, a challenge persists in determining the precise atomic structure at interfaces given that acquiring experimental data on interfacial morphologies often proves elusive[26]. Direct observation of interfacial atomic structures using high-resolution TEM is challenging, primarily because of the limited reflections from lattice planes which make it difficult to ascertain actual atom positions[27], and high-resolution observations of these interfacial geometries remain a formidable task[28, 29]. As a result, the detailed atomic structure and compositional data of interfaces required for precise first-principles calculations of their electronic properties is often lacking. This underscores the importance of a comprehensive theoretical investigation and systematical exploration of various interfacial morphologies, accounting for different degrees of lattice mismatch, and to analyze their impact on the electronic and band bending properties of heterostructures.

Theoretical calculations for a long time have been employed to study the electronic properties of interfaces. While earlier methods leaned on more basic techniques such as effective dipole models[30], tight-binding schemes[31] or empirical rules[32], in the recent few decades DFT has been utilized as an advanced computational to study band offset and interfacial dipole in heterostructures[33-36]. Despite the valuable contributions of DFT studies in this area, several limitations persist. The majority of investigations lack precise predictions due to the absence of explicit heterostructure calculations[37-47]. Additionally, many studies tend to focus on arbitrary surface orientations of the individual components rather than considering the most dominant orientations from experimental observations, and more importantly, arbitrarily chosen interfacial morphologies, which may not accurately sketch the full picture of all the possible configurations and their effect on band bending[15, 48-53].

In this study, two well-known photocatalysts $TiO_2$ and $Cu_2O$ have been selected to study the morphological effect on their interfacial electronic properties. $TiO_2$ —a prominent photocatalyst due to its remarkable photocatalytic properties[54-58]— has been frequently paired with $Cu_2O$ — an abundant, low-cost and well-studied effective photocatalyst[59]— demonstrating enhanced charge transfer rates and reduced electron-hole recombination[37, 40, 54]. This heterostructure, mitigates the main drawback of TiO$_2$ which is its large band gap (~3.2 eV[60]), by pairing it with Cu$_2$O, a lower band gap material (~2.2 eV[61]). The incorporation of Cu$_2$O facilitates the extension of light absorption into the visible spectrum, thereby optimizing the use of solar energy[54]. Furthermore, both materials have appropriate band edges with respect to the redox potential of many pollutants[62, 63], facilitating the requisite redox reactions by ensuring that the photogenerated electrons in the conduction band (CB) have adequate energy for reduction reactions, while the holes in the valence band (VB) possess sufficient energy for oxidation reactions. Numerous experimental studies have been conducted to examine the interfacial band edges of Cu$_2$O/TiO$_2$. While the majority of investigations have reported a staggered (type II or Z-scheme) heterojunction featuring Cu$_2$O band edges situated above the TiO$_2$ band edges[47, 64-80], a limited number of studies have identified a straddling (type I) heterojunction with the VB and CB of Cu$_2$O positioned within the band gap of TiO$_2$[54, 81]. This has led to a debate regarding the nature of the heterostructure between type II, Z-scheme, and type I configurations. Despite a great number of experimental studies, theoretical studies on Cu$_2$O/TiO$_2$ remain scarce[82-84]. Although experimental research has provided valuable insights into the interfacial properties of Cu$_2$O/TiO$_2$, a more comprehensive understanding of the electronic properties at the interface is still needed, particularly from a theoretical standpoint. In this study, we mapped all the possible interfacial morphologies of Cu$_2$O/TiO$_2$ and by assessing the chemistry of interfacial bonding, quantifying the strain magnitude, and considering the size of the system in relation to the computational feasibility, we identified eight distinct interfacial morphologies to study their band bending properties. Despite the variation in the morphology of the interfaces and the level of strain, we observed similar electronic properties and band bending behavior across all the heterostructures.

**Computational Details**

All calculations were carried out based on the framework of generalized Kohn-Sham scheme[85, 86] as implemented in the Vienna ab initio simulation package (VASP)[87-89], employing PBE[90, 91] and Heyd−Scuseria−Ernzerhof (HSE06) hybrid functional[92-94]. The electron–core interactions have been described by means of the projected augmented wave (PAW) method [89, 95]. Soft PAW potentials were used for Cu, O, and Ti atoms. The electronic wave functions were expanded in plane waves with an energy cutoff of 550 eV. The Brillouin zone was sampled using the Monkhorst–Pack special k-point mesh[96]. The convergence threshold for total energy self-consistency was kept at $10^{-6}$ eV. The atomic coordinates and unit cell parameters have been optimized using the conjugate gradient method until the force on each atom was less than 0.01 eV Å$^{-1}$. TiO$_2$ in its anatase phase has been used because of its superior photocatalytic activity.

The crystal configuration of Cu$_2$O is defined by a cubic lattice structure, characterized by Pn$\bar{3}$m space group. Its unit cell contains four copper (Cu) and two oxygen (O) atoms. The crystal configuration of TiO$_2$ (titanium dioxide) in its anatase form is defined by a tetragonal lattice structure, characterized by the I41/amd space group. Its unit cell contains four titanium (Ti) and eight O atoms. The computed lattice parameters using PBE functional for Cu$_2$O are $a_{Cu_2O} = 4.267$ Å , and for TiO$_2$ are $a_{TiO_2} = 3.807$ Å, $c_{TiO_2} = 9.707$ Å, in good agreement with the experimental findings of $a_{Cu_2O}^{exp} \sim 4.269$ Å[97, 98], $a_{TiO_2}^{exp} \sim 3.78$ Å[99-101], and $c_{TiO_2}^{exp} \sim 9.50$ Å[99-101].

In order to design a realistic model system of the Cu$_2$O/TiO$_2$ heterojunction, we have considered the Cu$_2$O (111) and the TiO$_2$-anatase (101) nonpolar surfaces, being the most stable and dominant terminations across diverse morphologies and fabrication techniques[37, 43, 54, 102-104]. The surfaces were modeled as two-dimensional periodic slabs, with a vacuum layer separating the periodic images in the z-direction. A vacuum region of at 20 Å was tested to be sufficient to avoid the superficial interactions between the periodic slabs. The in-plane lattice constants are fixed to the optimized bulk values and only the internal coordinates are relaxed. To characterize the surfaces, the surface energy ($\gamma$) as a measure of the thermodynamic stability has been calculated through the following expression:

$$\gamma = \frac{E(n) - nE_{\text{bulk}}}{2A}, \tag{1}$$

where $E(n)$ is the energy of the slab containing $n$ layers, $E_{\text{bulk}}$ the energy of the bulk, and $A$ the area of one side of the slab.

The specific adhesive energy, a measure of the energy gained once the interface boundary between two surfaces ($s1$ and $s2$) is formed, is given by:

$$\beta_{s1/s2} = \frac{E_{s1} + E_{s2} - E_{s1/s2}}{A}, \tag{2}$$

where $E_{s1}$ and $E_{s2}$ are total energies of the respective slabs and $E_{s1/s2}$ is the final interface energy. The specific interface energy, defined as the excess energy resulting from the energy balance described by the Dupré's relation[105], is given by:

$$\gamma_{s1/s2} = \gamma_{s1} + \gamma_{s2} - \beta_{s1/s2}, \tag{3}$$

where $\gamma_{s1}$ and $\gamma_{s2}$ are surface energies of the respective slabs forming the interface, and $\beta_{s1/s2}$ the adhesion energy defined earlier.

The planar and macroscopic averaged potentials as well as the charge density differences were computed using the VASPKIT post-processing code[106]. Graphical drawings were produced using VESTA[107] and OVITO[108].

The vertical ionization potential (IP) is calculated using a bulk-based definition, via the electrostatic alignment between the surface and the bulk as follows[109]:

$$\varepsilon_{IP} = \Delta\varepsilon_{\text{vac-ref}} - \Delta\varepsilon_{\text{VBM-ref}},$$

where $\Delta\varepsilon_{\text{vac-ref}}$ is the difference between the electrostatic potential in the vacuum region and the bulk-like reference level in the slab (the 1s states of Cu and Ti in the middle of the Cu$_2$O(111) and TiO$_2$(101) slabs, respectively). The second term, $\Delta\varepsilon_{\text{VBM-ref}}$, is the difference in eigenvalue energy between the valence band maximum (VBM) and reference level from bulk calculations. The electron affinity (EA) is obtained by subtracting the obtained bang gap value from the ionization potential.

The macroscopic average of the electrostatic potential along the non-periodic direction of the interfaces was computed using the MacroDensity package[110].

The heterostructures with different degrees of lattice mismatch (strain) were generated using the QuantumATK software developed by Synopsys[111] which provides a user-friendly setting to adjust the lattice vectors and align the crystal structures to create the desired heterostructures. In generating the interfaces, Cu$_2$O was subjected to strain (treated as a film), a consequence of its deposition as the top layer in the Cu$_2$O/TiO$_2$ fabrication process[112-114].

**Results and Discussion**

**The surfaces**. Determining suitable surface orientations for the construction of a heterostructure poses a challenge due to its impact on band bending characteristics. This challenge arises from the involvement of surface dipoles in the determination of the VB and CB edges, which are critical factors in establishing interfacial band alignment. These band positions are intrinsically influenced by the surface orientation, composition, atomistic and electronic structure, all of which collectively contribute to the surface dipole effects[115-118]. From a computational standpoint, a key constraints in selecting the right surfaces is the assessment of surface polarity, as polar surfaces inherently exhibit a notable electrostatic instability[119]. This instability arises from the presence of macroscopic dipoles oriented perpendicular to the surface within each unit cell that accumulate, necessitating the introduction of compensating charges to neutralize these dipoles. Achieving such compensation in practice involves extensive surface modifications, including significant adjustments in stoichiometry, faceting, spontaneous desorption of atoms, large-cell reconstruction due to the ordering of surface vacancies, among other complex processes[120, 121].

Experimentally, the primary surfaces in heterostructures are identified through high-resolution TEM analysis. This method unveils predominant lattice spacings, facilitating the determination of the corresponding surfaces for each constituent material within the heterostructure. In the specific case of the $Cu_2O/TiO_2$ heterojunction, surfaces (111) and (101) have consistently emerged as the dominant surfaces for $Cu_2O$ and $TiO_2$, respectively, across diverse morphologies and fabrication techniques[37, 43, 54, 102-104]. Notably, these surfaces exhibit non-polar characteristics, showing no electrostatic instabilities. Therefore, in the current study, surfaces (111) and (101) are selected for $Cu_2O$ and $TiO_2$, respectively.

The initial geometry of the slabs for a 4-bilayer $TiO_2$ and 6-trilayer $Cu_2O$ are shown in **Figure S1**. The thickness of the surfaces has been optimized with respect to the surface energy (**Equation 1**). The results using PBE functional are presented in **Figure S2** which indicate that the surface energy quickly converged to a value of 1.13 $J/m^2$ for a 4-bilayer $TiO_2$ and 0.77 $J/m^2$ for a 6-trilayer $Cu_2O$, in agreement with the earlier studies[122, 123].

**Interfaces.**

**Geometry.** To construct the interfaces, we have confined this study to epitaxial heterostructures, with lattice mismatch values predominantly confined within a 10% threshold[124, 125]. It is well-established that an escalated degree of lattice mismatch can lead to the formation of defects in the heterostructure, such as dislocations, threading dislocations, misfit dislocations, and cracks[126]. Severe lattice mismatch will cause dislocations at the interface and results in electrical defects such as interface traps[127]. These defects can significantly degrade the electronic and optical properties of the heterostructure, which can limit their practical applications. **Figure 1** shows a map of all the possible heterostructure of $Cu_2O(111)/TiO_2(101)$ as a function of the mean absolute strain (lattice mismatch) −which is defined as $\frac{a_{TiO_2}-a_{Cu_2O}}{a_{TiO_2}}$, where $a$ is lattice constant− and number of atoms in the unit cell of the heterostructures.

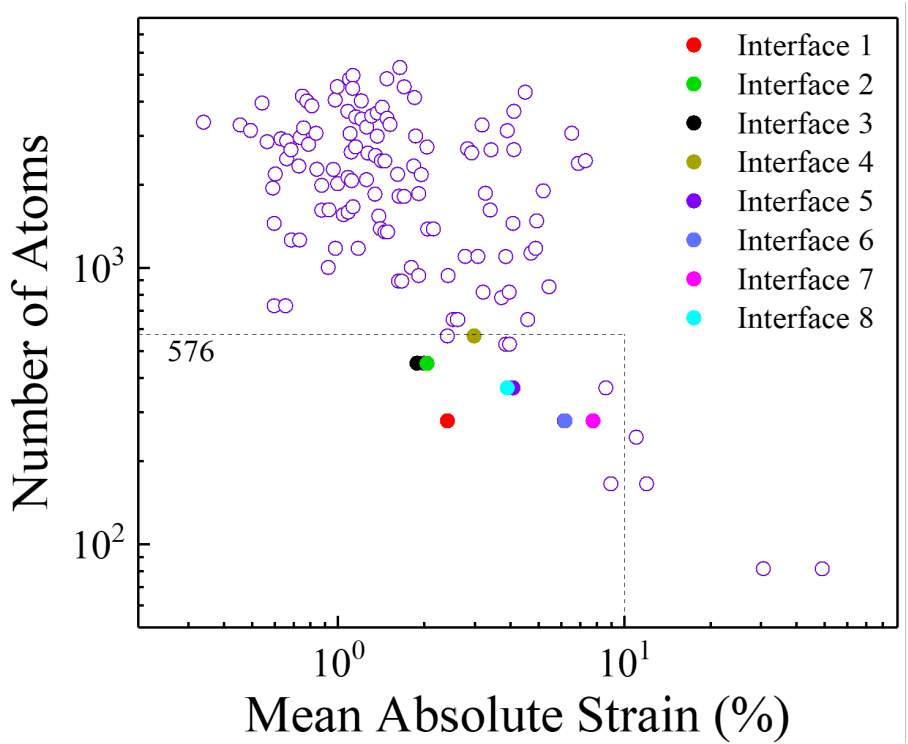

*Figure 1. Map of all the possible heterostructures of $Cu_2O(111)/TiO_2(101)$ as a function of lattice mismatch (as defined in the text), and total number of atoms in the unit cell. The eight selected interfaces, confined within 576-atom limit, and 10% mean absolute strain, are highlighted in different colors.*

In the current study, a systematic approach was employed to select the interfaces, ensuring that the strain remained within a limit of 10%, while also maintain computational feasibility (total number of atoms ≤ 576). As shown in **Figure 1**, this criterion resulted in eight distinct interfaces for subsequent analysis. A detailed numerical analysis of the strain present at the selected interfaces is outlined in **Table 1**.

*Table 1. Details of the strain matrices in the supercell of the selected interfaces and corresponding adhesive and interface energies, where $\epsilon_{xx}$ and $\epsilon_{yy}$ are the normal strains along the x and y directions and $\epsilon_{xy}$ represents the shear strain in the x-y plane.*

| Interface | $\epsilon_{xx}$(%) | $\epsilon_{yy}$(%) | $\epsilon_{xy}$(%) | Mean absolute strain (%) | Adhesive en. β (J/m²) | Interface en. γ (J/m²) |
|---|---|---|---|---|---|---|
| 1 | 5.87 | -1.39 | 0.00 | 2.42 | 2.29 | -0.13 |
| 2 | 4.59 | -1.39 | 0.00 | 2.03 | 2.94 | -0.95 |
| 3 | 3.12 | 0.01 | -2.55 | 1.89 | 1.41 | 0.48 |
| 4 | -5.15 | -2.14 | -1.66 | 2.98 | 2.03 | 0.04 |
| 5 | -5.13 | 4.36 | -2.74 | 4.07 | 1.48 | 0.53 |
| 6 | -9.02 | 2.02 | 7.52 | 6.19 | 1.90 | 1.16 |
| 7 | -14.60 | 8.69 | 0.00 | 7.76 | 1.43 | 1.82 |
| 8 | 5.05 | -5.76 | -0.88 | 3.90 | 1.27 | 0.48 |

From the geometrical standpoint, the main connecting point between the two materials are the outermost oxygen atoms, which is a standard bridging mechanism for metal oxides. On the TiO$_2$(101) surface, there are the undercoordinated topmost surface O atoms and subsurface Ti atoms, while on the Cu$_2$O(111) surface there are subsurface undercoordinated Cu atoms present in the same plane with coordinatively saturated Cu atoms linking two O atoms, while the topmost surface O atoms undercoordinated. However, due to the approximately 2:1 ratio of available Ti to Cu atoms (per unit cell or comparable surface area) as well as $O_{TiO_2}^{top}$ to $O_{Cu_2O}^{top}$ atoms, it is to be expected that a certain number of bonds will remain undercoordinated.

Upon relaxation, the TiO$_2$ topmost 2-fold coordinated O atoms formed a bond at the interface with the singly coordinated Cu atoms on the Cu$_2$O side and the 3-fold coordinated O atoms from the Cu$_2$O side bonded with the 5-fold undercoordinated titanium atoms from TiO$_2$. However, depending on the amount of lateral and shear strain present in the initial model, the number of formed bonds differs greatly amongst the eight chosen interfaces. This can be observed in the calculated adhesive energy, listed in **Table 1**, which is a direct quantity (at the computational level) allowing one to evaluate the probably of observing the corresponding epitaxial interface. From the computed values, it is notable that not all interfacial structures result in an equal energy gain once the interface is formed.

The energetically most stable interface is found to be interface 2, which is shown in **Figure 2**. This interface has the lowest amount of lateral strain (present in the film), while at the same time undergoing no shear strain. In this configuration, all the O atoms from the substrate and film are coordinatively saturated upon relaxation (completion of their octet), while every other Ti atom remains undercoordinated.

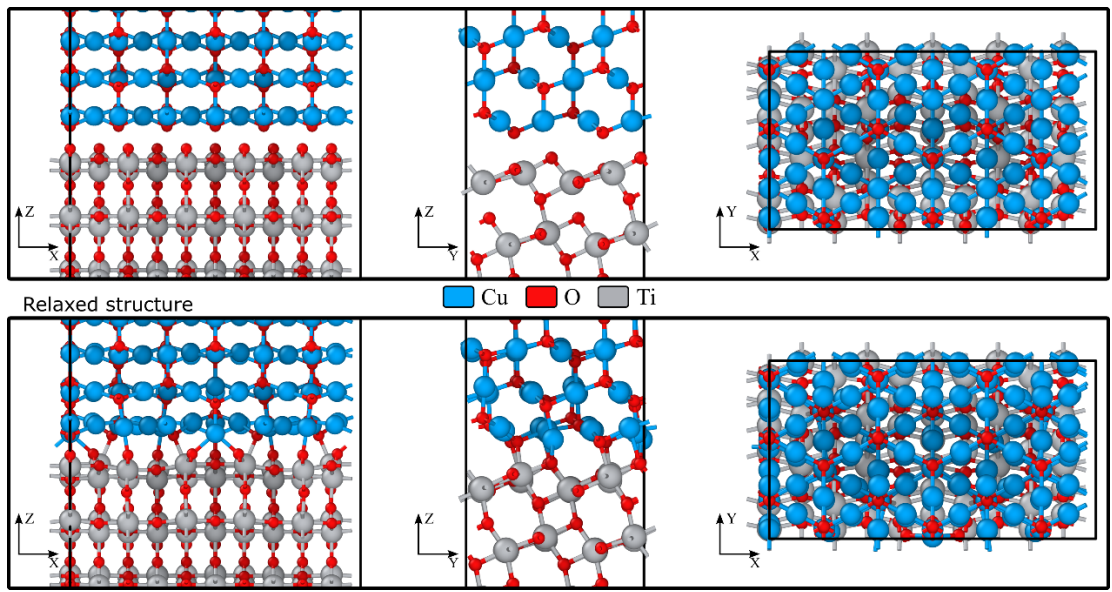

*Figure 2. Initial and relaxed atomic structure of the most stable $Cu_2O/TiO_2$ interface 2, viewed along all three crystallographic axes. Vacuum is present along the Z-axis but is omitted for clarity.*

The accurate assessment of band bending properties at interfaces requires a comprehensive understanding of the structural and electronic changes induced by the formation of the interface. The independent slab approximation only provides an incomplete representation of the band alignment, resulting in significant discrepancies when compared to experimental data[128]. Explicit interfacial relaxations can induce a shift in the band alignment as much as 130 meV highlighting the key role of the relaxation in the precise estimation of band alignment[129]. The relaxed structures of interfaces 1-8 are shown in **Figure S3**.

**Band alignment.** When two semiconductor materials form an interface, a discontinuity in their band edges occurs. In the computational evaluation of these relative band edges, a challenge arises due to the absence of an absolute reference energy in an infinite solid attributed to the long-range nature of the columbic interaction[130, 131]. To address this, the average electrostatic potential has been suggested as a consistent energy reference in such systems[132-134]. The method involves first determining the average of the electrostatic potential of the interface, followed by calculating the band edges of the individual slabs relative to their average electrostatic potential for the individual slabs. It is crucial to account for the strain induced by interfacial relaxations; therefore, the relaxed slab geometries should be utilized in these calculations without further relaxations[48].

**Functional.** Within the framework of DFT, generalized gradient approximation (GGA) functionals suffer from self-interaction errors which result in the systematic underestimation of band gaps and the overestimation of cohesive energies in materials[135]. To address these limitations, hybrid functionals, which integrate a mix of Fock exchange with semilocal exchange, are designed to mitigate this delocalization error[136]. The partial incorporation of the exact exchange from Hartree-Fock theory into these functionals helps to correct the self-interaction error, thereby providing a more accurate representation of electron correlation and localization. This adjustment is crucial for electronic properties as it leads to a more precise prediction of band gaps. However, hybrid functionals are computationally demanding and the relaxation of the heterostructures comprising hundreds of atoms becomes impractical. Consequently, a synergic approach that combines the computational efficiency of GGA functionals with the accuracy of hybrid functionals can be employed. This combined approach enables an accurate assessment of the band bending, at a reasonable computational cost. It consists of heterostructure relaxation via GGA functionals and assessment of band edge positions of isolated slabs via hybrid functionals.

In calculating the band alignment, since the average electrostatic potential is a function of the ground state charge density, and the difference in the distribution of electronic density obtained with different functionals which determines the Hartree potential, is small[137, 138], the potential lineup can be accurately determined via GGA[129, 135, 139-141] and many-body corrections[142, 143] or hybrid functionals such as PBE0 schemes[129] can barely affect the band potential lineup in accordance with findings for semiconductor-insulator interfaces[135, 141]. For instance, it has been shown that the potential alignments calculated within the GGA are in agreement with those calculated using HSE to within 50 meV, despite the GGA calculations being 10–100 times less computationally expensive[138]. In other hybrid functional study of Si and $TiO_2$ surfaces[144] and a self-consistent GW study of an Si/$SiO_2$ interface[135] have indeed shown that the changes in the averaged electrostatic potential from the semilocal values are less than 0.1 eV at these surfaces and interface. Moreover, the functional dependence of the formation energies of Si self-interstitials as a function of the electronic chemical potential has shown that taking the average electrostatic potential as the reference of the energy results in invariant features in the formation energy for any given value of chemical potential[137]. Thus, considering the accuracy of potential alignments via GGA, they can be combined with hybrid calculations of the isolated slabs' band structure toward accurate band alignments with a high computational efficiency[138].

While GGA calculations present a lower computational cost compared to more sophisticated methods, the computational demands still escalate significantly for large-scale systems. Therefore, it is imperative to employ the minimal slab thicknesses and thus lowest possible number of atoms in the calculations. This reduction in system size in justified by the localized nature of interfacial electronic properties at neutral interfaces[139]. The interfacial properties are usually confined within a few atomic units from the interface beyond where the bulk properties of charge density rapidly converge. To consider the smallest possible interface and verify the assessment of the slab thicknesses via surface energy (**Figure S2**), the average electrostatic potential of interface 1 is calculated in comparison to the independent slabs' approximation in **Figure 3**. Since $TiO_2$ is the substrate and does not undergo strain (only chemical alterations at the interface itself), it's electrostatic potential, as shown in **Figure 3a**, remains relatively unchanged when compared with the explicit interfaces depicted in **Figure 3c** and **3d**. In contrast, due to the imposed strain on $Cu_2O$, it is more influenced by the interfacial effects which only penetrate no more than two atomic layers from the interface as can be seen in **Figure 3c**. To validate the sufficiency of the 6-layer $Cu_2O$ slab, an 8-layer slab is also examined in **Figure 3d** which reveals a continuation of the bulk-like behavior within the central layers of the slab. This finding suggests that the 6-layer $Cu_2O$ slab adequately captures the intrinsic electrostatic characteristics of the bulk. Consequently, $Cu_2O$(6-layer)/$TiO_2$ (4-layer) model configuration is selected throughout this study.

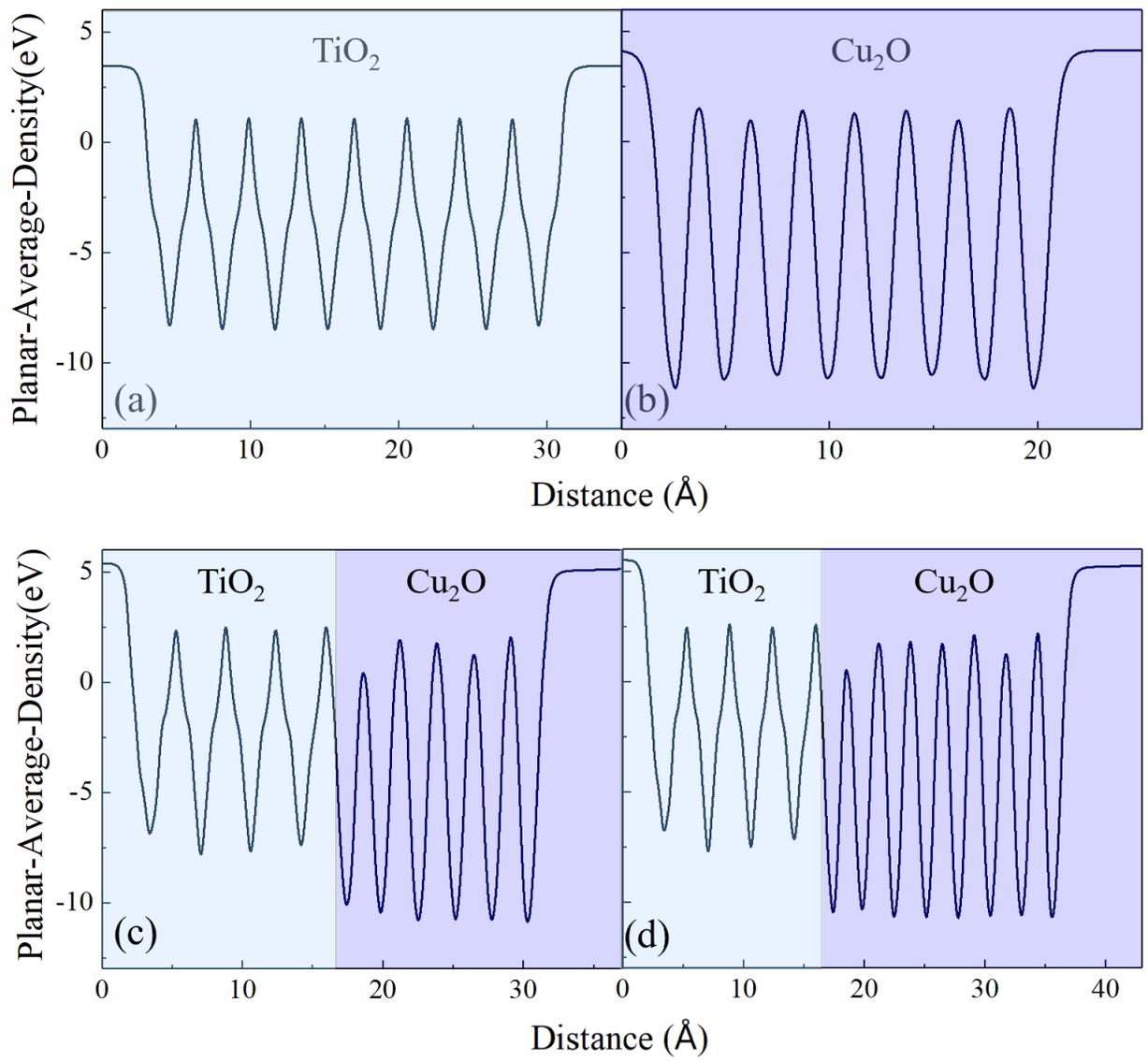

*Figure 3. Thickness assessment of interface 1 in relation to the thickness of the $Cu_2O$ through planer-averaged electrostatic potential: (a,b) independent slab approximation (before contact). (c) explicit interface of $Cu_2O/TiO_2$ with 6 layers of $Cu_2O$, and (d) 8 layers of $Cu_2O$.*

As aforementioned, an electrostatic potential approach has been employed to assess the band bending of the various interfaces. This method entails calculating the electrostatic potential, averaged across the x-y plane. The resulting data for interface 2 is depicted in **Figure 4**, where the planar-averaged electrostatic potential is plotted as a function of the distance along the z-axis, normal to the interface plane. Notably, this profile shows distinct minima corresponding to atomic planes, indicating areas with higher electron density. In the section of the interface composed of $TiO_2$ **Figure 4a**, notable deviations are observed in the potential profile. These deviations are attributed to the displacement of O atoms relative to Ti-containing planes, a structural feature inherent to the anatase phase of $TiO_2$. This atomic arrangement results in characteristic 'shoulders' in the potential curve, observable throughout the entire extent of the anatase layer.

When determining an appropriate reference point on the electrostatic potential curve, concerns arise regarding the precision of planar-averaged potentials at planes containing atoms. The main challenge comes from the necessity to integrate the electrostatic potential in proximity to atomic positions, where the potential changes very quickly and sharply. These steep variations in potential make it difficult to compute the potential accurately, particularly over confined regions such as within a single lattice period. To address this issue, as outline by Conesa[134], in this study, instead of relying on the planar-averaged potential at the atom-containing planes, we utilized the energy value at the point of zero slope (termed $E_{top}$). These zero-slope points typically occur near the midpoint between successive atomic planes. This method is considered more reliable for determining the reference value for electronic levels. The analysis reveals that these zero slope values display oscillations within each slab near the interface, but they stabilize to a constant value towards the center of each slab. The resultant values for interface 2 are $E_{top}$ = 2.11 eV and 1.79 eV for the $TiO_2$ and $Cu_2O$ regions, respectively.

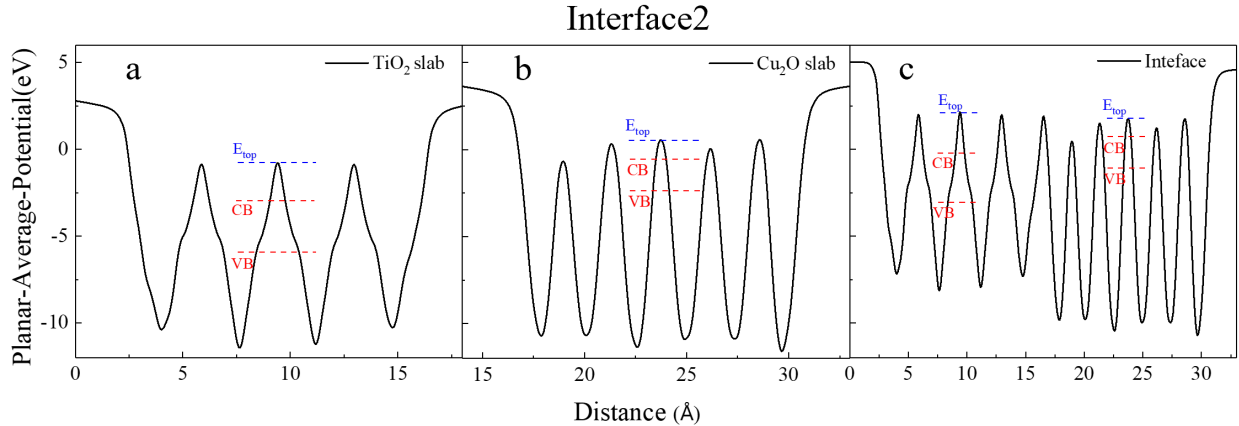

*Figure 4. Planar-averaged electrostatic potentials in the x-y plane of (a,b) the isolated slabs with the interface geometries, and (c) of interface 2 as a function of the distance along the z-direction, normal to the interface. $E_{top}$ denotes the reference energy, and the CB and VB edges in the interface are aligned based on their respective values to the $E_{top}$ in the isolated slabs' calculations.*

After establishing the reference energies ($E_{top}$), the next step is determining the band edges of the individual slabs relative to these reference energies. To achieve this, the planar-averaged electrostatic potential was computed for the individual $TiO_2$ and $Cu_2O$ slabs. However, it is important to note that the slabs acting as films are distorted due to epitaxial strain and to accurately represent the electronic properties under these conditions, it is imperative to calculate the electrostatic potentials for the distorted structures. Thus, hybrid functional calculations were performed, considering the distorted slab structures as they exist at the interface without any further relaxations. **Figure 4a,b** show the results of the hybrid calculation for the individual $TiO_2$ and $Cu_2O$, respectively where the relative positions of the VB and CB edges are shown with respect to the $E_{top}$ for each slab. With the band edges of both materials positioned relative to the reference energy, the final band offsets are conclusively determined and depicted in **Figure 4c**. The findings suggest a staggered-type alignment at the interface wherein the band edges of $Cu_2O$ are positioned at a higher energy level compared to those of $TiO_2$. Similar results are obtained for the other interfaces and the details are presented in **Figures S4-10**. The band alignment results for all the interfaces are summarized in **Figure 4**. This figure highlights the variations in the positions of the VB and CB edges across different interfaces as a result of their different interfacial morphologies. Despite these variations, a key observation is the consistent presence of a staggered-type alignment across all the interfaces. The values of the band offsets as noted in the graph, are comparable to the experimental values of $\Delta E_v =1.93$ eV[145], 1.71 eV[146], $\Delta E_c =0.81$ eV[145], 0.88 eV[146]. This uniformity in the alignment type, despite differences in the specific energy levels of the band edges, underscores a fundamental characteristic of these heterojunctions.

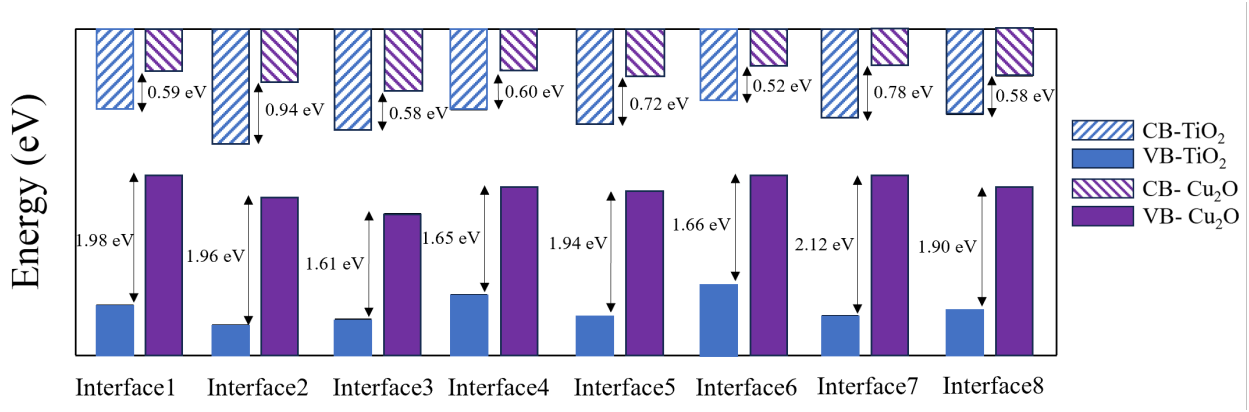

*Figure 5. Summary of the band alignment results for interfaces 1-8. The band offsets, $\Delta E_v$ and $\Delta E_c$, are indicated for each interface.*

For comparison, we computed the valence and conduction band offsets in the independent compound approximation and through the averaged electrostatic potential method (results listed in **Table S1** and **Table S2**). The computed ionization potential (IP) and electron affinity (EA) of $TiO_2$ and $Cu_2O$ compare well with experimental values, however, the band offsets are overestimated in this approach. When employing the macroscopic averaging technique of the interface electrostatic potential to obtain the offset in the potential across the interface, the results (1.93 eV for VBO and 0.42 eV for CBO) compare much closer to the measured experimental values. This confirms once again how indispensable explicit modelling of interfacial structures is for accurate electronic band behavior at junctions.

While band alignment configurations provide valuable insights into the arrangement of band edges at the interface, they do not conclusively establish the charge transfer properties. Specifically, as illustrated in **Figure 6**, within a staggered-type alignment between two semiconductors (SC1 and SC2), there exist two possible configurations for charge transfer. **Figure 6a** depicts a type II configuration, wherein charge separation is achieved via the antiparallel transfer of photoexcited electrons and holes. in the opposite direction. Whereas **Figure 6b** depicts a scenario were the photoexcited electrons in SC2 recombine with the holes in SC1 through a Z-scheme mechanism.

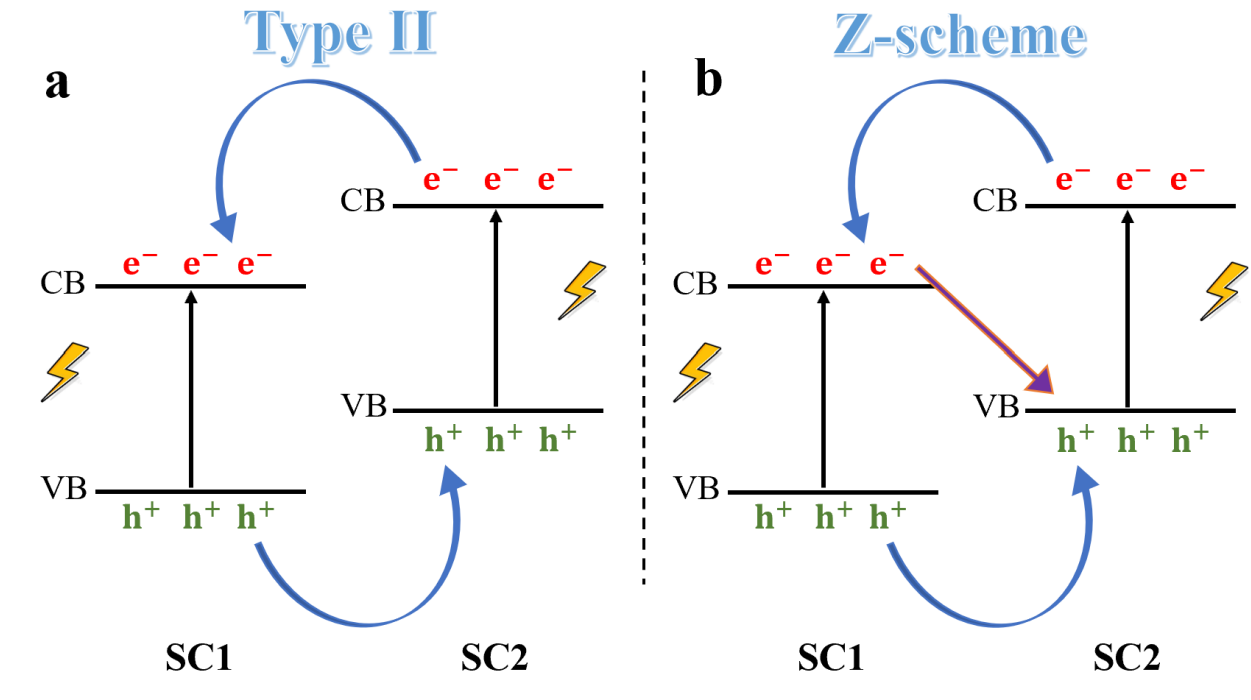

*Figure 6. Schematic illustration of charge transfer mechanism after light illumination. (a) type II where electrons and holes transfer in an antiparallel way, and (b) Z-scheme where after the antiparallel charge transfer, electrons in SC1 combine with the holes in SC2 (SC stands for semiconductor).*

Hence, to differentiate between a type II and Z-scheme charge transfer mechanisms, it is essential to quantify the charge density difference at the interfaces, defines as

$$\Delta \rho = \rho_{interface} - \rho_{TiO2} - \rho_{Cu2O}, \tag{2}$$

where $\Delta\rho$ is the charge density difference at the interface, $\rho_{interface}$ is the charge density of the interface, and $\rho_{TiO2}$ and $\rho_{Cu2O}$ are the charge densities of the corresponding slabs before contact. **Figure 7** shows $\Delta\rho$ for interface 2 where the dashed line represents the midpoint layer of the interface. The figure suggests a directional charge transfer towards the interface from both the Cu$_2$O and TiO$_2$ layers. However, a closer inspection of the structure reveals that the electrons are transferred from the transition metal towards the oxygen. That is nothing else than the bond formation process confirmed which was discussed when the adhesion energies were discussed. The two charge depletion regions in **Figure 7** are approximately half the value of the main charge accumulation region at the interface. In addition, we computed the electron dipole moment at the interface. A maximal value of 0.1 D was obtained, arising from the disparity between the number of metal ions present to form bonds at the interface (see discussion on structure at the interface), which offers space for band bending to occur.

The charge density difference for all other interfaces is presented in **Figures. S11-17**. A consistent pattern of charge transfer between the transition metal and the interface oxygen is found across all interfaces. However, one should treat these results with care as they have been computed with a semi-local functional, thus not entirely circumventing the well-known charge delocalization error for *d*-electron bearing transition metals, which can affect the value of the interface dipole and corresponding charge transfer.

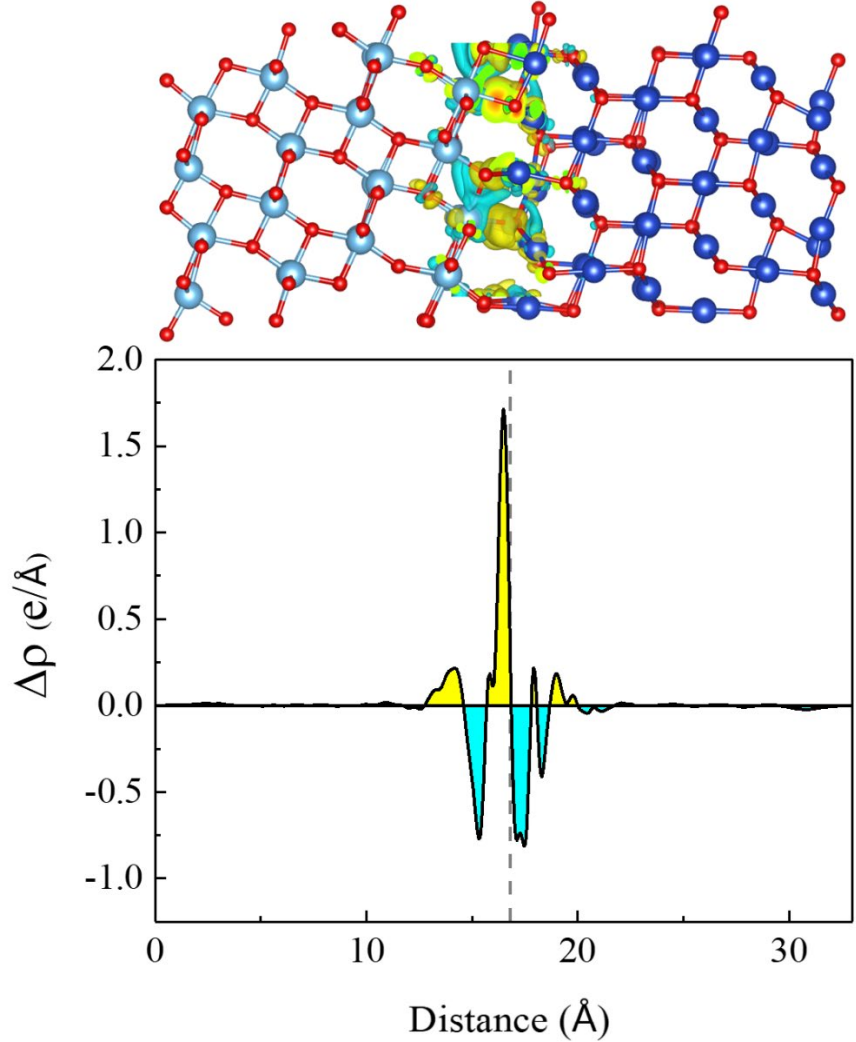

*Figure 7. Charge density difference of interface 2. Yellow: charge accumulation. Cyan: charge depletion. The dashed line represents the midpoint layer of the interface. Isosurface value is set to 0.0015 for the orbital visualization.*

To further verify the charge transfer direction, fermi energies ($E_F$) are evaluated before and after contact. When two semiconductors with different fermi energies come into contact, charge transfer occurs from the material with the higher $E_F$ to the one with the lower $E_F$ until the fermi levels equilibrate. **Figure 8c** shows the $E_F$ of interface 2 in comparison with the $E_F$ of the individual slabs before contact (**Figure 8a,b**). It is evident that $Cu_2O$ has a higher $E_F$ compared to $TiO_2$. Consequently, upon contact in the interface, charges transfer from $Cu_2O$ to $TiO_2$ until the fermi levels reach equilibrium. This charge transfer direction is consistent with the charge depletion observed in $Cu_2O$ and charge accumulation in $TiO_2$, as seen in **Figure 7**. The fermi energy assessment for other interfaces is shown in **Figures. S18-24**. The findings suggest the same charge transfer mechanism across all the interfaces.

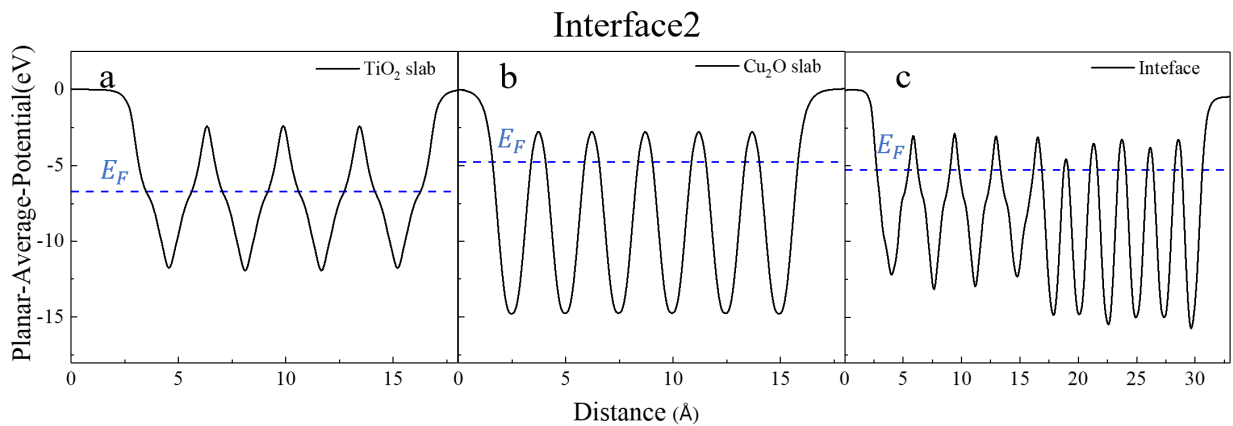

*Figure 8. Fermi energy assessment of interface 2 (a,b) before and (c) after contact*

The comprehensive analysis of charge transfer mechanisms in **Figure 4, 5, 7, and 8**, leads to the proposal of distinct behavior under visible and ultraviolet light illumination, specifically focusing on the photocatalytic behavior of the $Cu_2O/TiO_2$ interface. As shown in **Figure 9a**, under visible light, only $Cu_2O$ would be photoexcited due to its suitable band gap. This could potentially result in the generation of photoexcited electrons and holes within $Cu_2O$. However, the band bending configuration at the interface, given the charge depletion and accumulation in $Cu_2O$ and $TiO_2$, respectively, causes the $Cu_2O$ bands to bend upward and $TiO_2$ to bend downward, forming a Z-scheme band bending configuration. In this configuration, the photoexcited electrons in $Cu_2O$ would encounter an energy barrier that hinders their transfer to $TiO_2$. Consequently, these photoexcited electrons and holes are unable to separate efficiently, leading to their recombination within the $Cu_2O$ material. This recombination typically occurs on a very fast timescale (femtosecond to picosecond), which is much quicker than the timescale of chemical reactions (milliseconds)[147]. As a result, these carriers recombine before they can participate in photocatalytic reactions, leading to reduced photocatalytic activity under visible light.

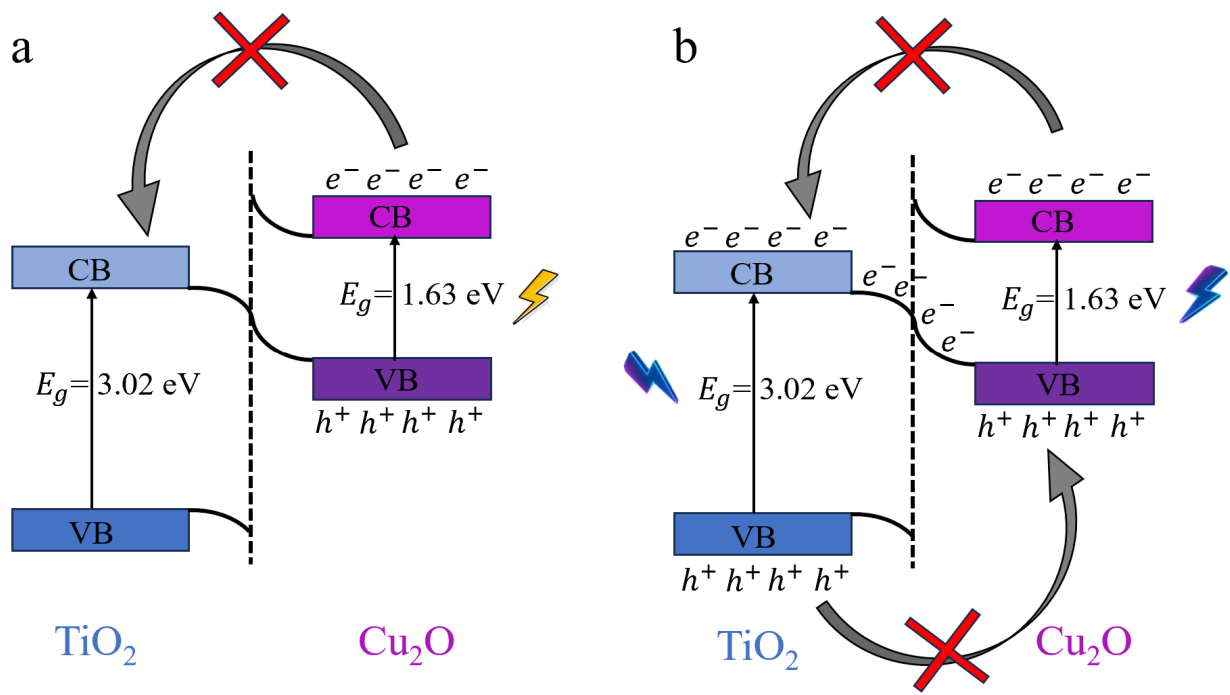

*Figure 9. The proposed band bending configuration for $Cu_2O/TiO_2$ interfaces under a) visible and b) ultraviolet light illumination.*

In contrast, under ultraviolet or violet light illumination, both $Cu_2O$ and $TiO_2$ are excited. In this scenario, the Z-scheme band bending plays a pivotal role in dictating the charge transfer dynamics. Specifically, the photoexcited electrons from $TiO_2$ recombine with the photoexcited holes in $Cu_2O$. This recombination process effectively separates the photoexcited holes in $TiO_2$ from the photoexcited electrons in $Cu_2O$. The separated charge carriers – photoexcited holes on $TiO_2$ and photoexcited electrons on $Cu_2O$ – are now available to participate in photocatalytic chemical reactions. This separation enhances the photocatalytic efficiency under ultraviolet light, as the carriers are able to participate in redox reactions. This observed Z-scheme band alignment configuration is consistent with numerous experimental investigations that have analyzed similar crystal facets, specifically $TiO_2$ (101) and $Cu_2O$ (111)[78, 114, 148-151].

**Conclusion**

In conclusion, this study delves into the intricate interplay between interfacial morphologies, band bending effects, and charge transfer dynamics in anatase $Cu_2O/TiO_2$ epitaxial heterostructures for photocatalytic applications. Through a systematic exploration, starting from well-defined $TiO_2$ (101) and $Cu_2O$ (111) slab structures, explicit epitaxial heterojunctions where $TiO_2$ serves as a substrate for a $Cu_2O$ film were created. Out of the obtained structures, those with a lattice mismatch of less than 10% were selected for further optimization. Eight distinct interfaces were identified, all with a consistent staggered-type band alignment, providing valuable insights into the fundamental characteristics of these heterojunctions.

Despite variations in band edge positions, a systematic charge transfer from $Cu_2O$ to $TiO_2$ was observed across all interfaces. The proposed band bending can help explain the nuances of charge separation at the $Cu_2O/TiO_2$ interface evidenced in earlier experiments. The observed Z-scheme band alignment under ultraviolet light aligns with experimental investigations, highlighting its relevance to enhancing photocatalytic efficiency. This theoretical investigation contributes to the understanding of the structural and electronic factors influencing the photocatalytic behavior of heterostructures, paving the way for the rational design and optimization of photocatalytic materials for environmental remediation applications.

It is essential to recognize that the experimental processes inherently introduce defects and impurities which can significantly influence the band bending properties of interfaces. Future research should therefore be directed towards a detailed investigation of how these unavoidable imperfections affect the electronic structure of the interfaces. This entails a systematic study of the type and concentrations of defects and impurities, as well as their spatial distribution within the material. Additionally, it is crucial to explore the mechanisms through which these defects and impurities interact with materials' electronic states. Such studies are vital for advancing our theoretical understanding of the interfaces as well as providing directions for experimental studies.

**Acknowledgments**

A. Ž. and N. H. d. L. acknowledge the NWO ECHO grant (712.018.005) for funding.

**Supplementary information**

**Interfacial Morphology and Band Bending in anatase $TiO_2$/$Cu_2O$ Heterostructures**

Mona Asadinamin,[1] Aleksandar Živkovic,[2,3] Nora H. De Leeuw,[2] and Steven P. Lewis[1]

[1] *Department of Physics and Astronomy, University of Georgia, Athens, Georgia 30602, US*

[2] *Department of Earth Sciences, Utrecht University, Princetonlaan 8a, 3548CB Utrecht, The Netherlands*

[3] *Institut für Theoretische Physik und Astrophysik, Christian-Albrechts-Universität zu Kiel, Leibnizstraße 15, 24118 Kiel, Germany*

*Corresponding author:* a.zivkovic@uu.nl, splewis@uga.edu

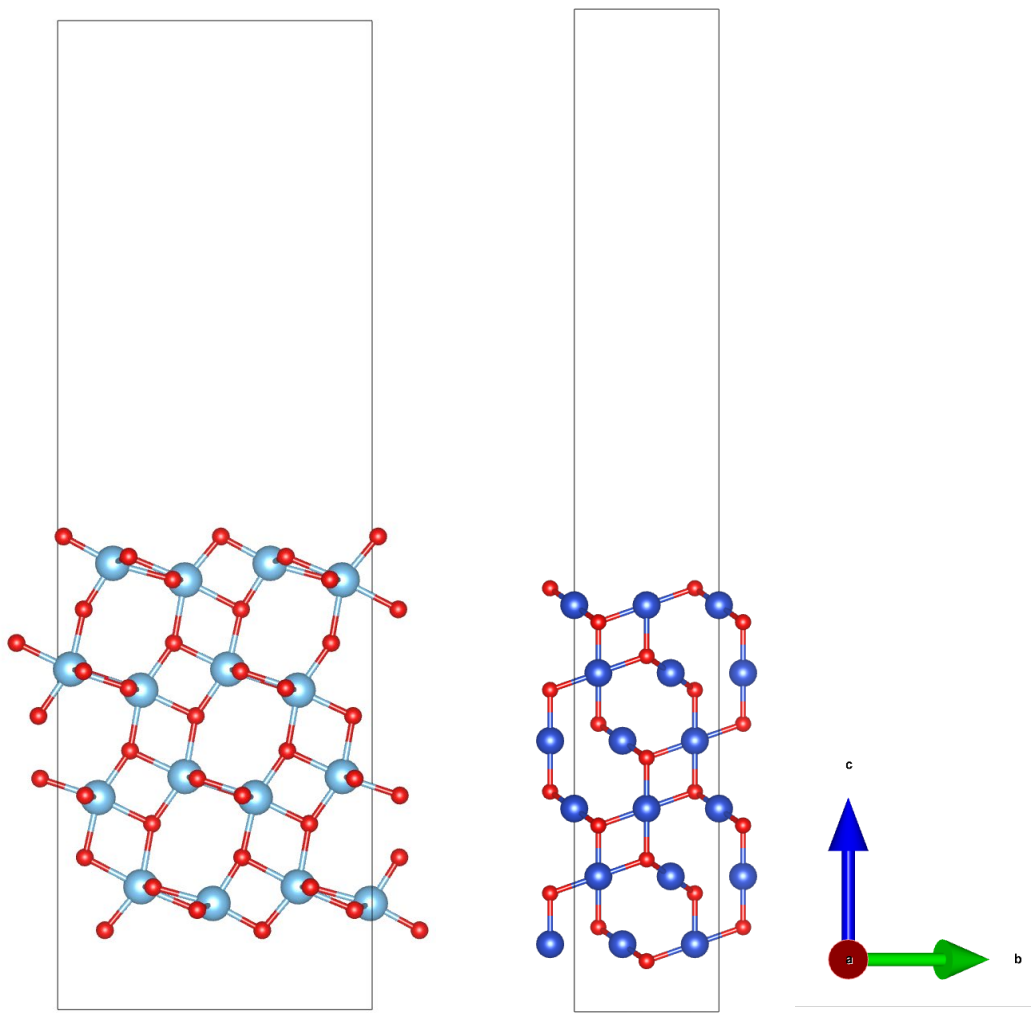

*Figure 10. Side view of the slab morphologies of the 4-bilayer $TiO_2$ (101) surface (left) and 6-trilayer $Cu_2O$ (111) surface (right). Ti: light blue, oxygen: red, and cu: dark blue. The coordinate system is shown by the blue and green arrows where c indicates the non-periodic direction.*

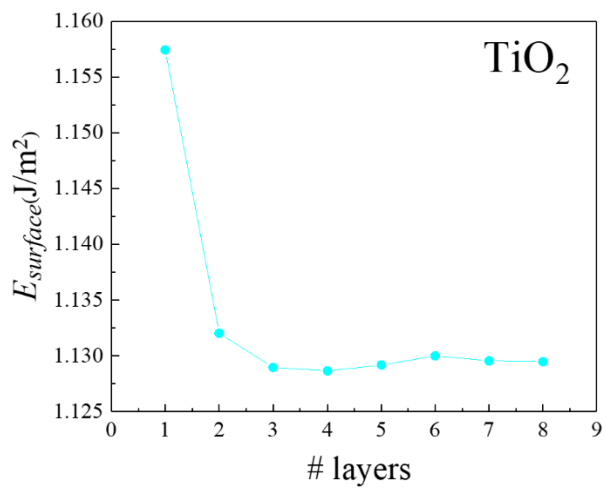 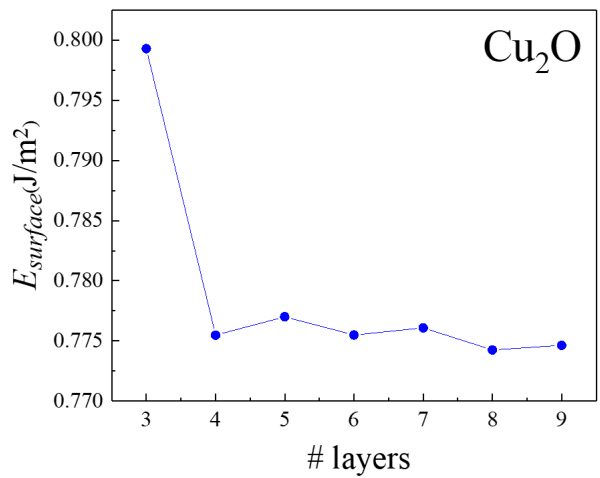

*Figure 11. Surface energies of the relaxed TiO$_2$ (left) and Cu$_2$O (right) slabs which quickly converged to a value of approximately 1.22 J/m2 for a 4-bilayer TiO$_2$ and 0.77 J/m2 for a 4-trilayer Cu$_2$O.*

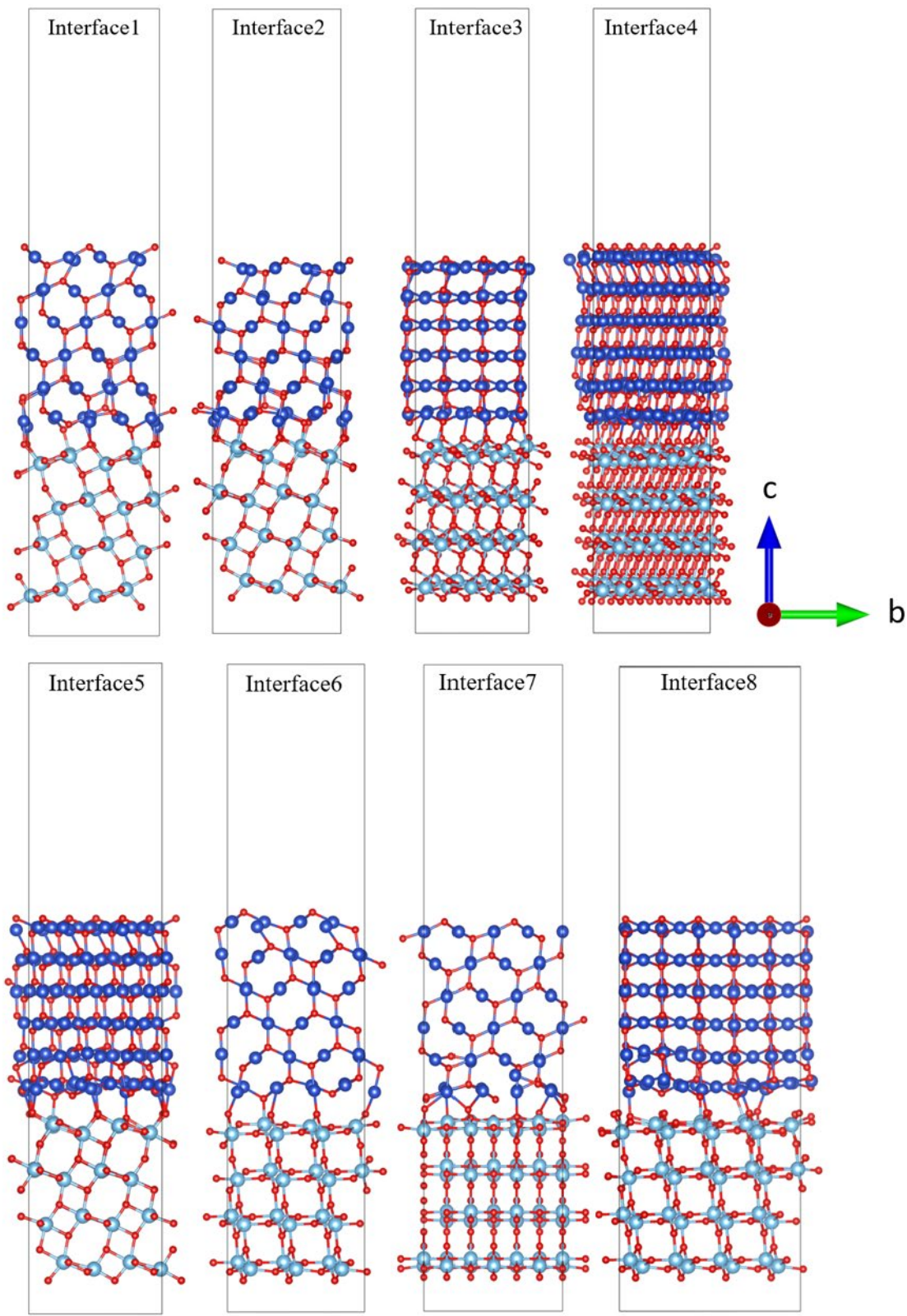

*Figure 12. Side view unit cell of the relaxed structure of interfaces 1-8. The green and blue arrows show the coordinate systems.*

*Table S2. Computed values of electronic band gap, ionization potential (IP), and electron affinity (EA) for Cu2O and TiO2.*

| Compound | XC | Kohn-Sham gap (eV) | IP (eV) | EA (eV) | Exp. (eV) |
|---|---|---|---|---|---|
| $Cu_2O$ | PBE | 0.46 | 4.64 | 4.18 | 5.0 eV to 4.0 eV (IP[1]) |
| | HSE06 | 1.94 | 5.81 | 3.87 | 3.20 eV (EA[2]) |
| $TiO_2$ | PBE | 1.94 | 7.00 | 5.03 | 7.96 (IP[3]) |
| | HSE06 | 3.34 | 8.84 | 5.30 | 5.1 to 5.3 eV (EA[4]) |

*Table S3. Computed valence band and conduction band offsets at the $Cu_2O$ / $TiO_2$ interface using two different approaches, the independent compounds alignment and the alignment based on an explicitly modelled interfacial structure.*

| System | XC | Valence band offset (independent compound alignment) (eV) | Conduction band offset (independent compound alignment) (eV) | Valence band offset (explicit interface) (eV) | Conduction band offset (explicit interface) (eV) |
|---|---|---|---|---|---|
| $Cu_2O(111)$ / $TiO_2(101)$ | PBE | 2.35 | 0.84 | 1.93 | 0.42 |
| | HSE06 | 2.83 | 1.43 | 1.92[1] | 0.52 |

---

[1] The potential offset at the interface was taken from PBE calculations, assuming transferability of values as outlined in the work of (5) Conesa, J. C. Modeling with Hybrid Density Functional Theory the Electronic Band Alignment at the Zinc Oxide–Anatase Interface. *The Journal of Physical Chemistry C* **2012**, *116* (35), 18884-18890. DOI: 10.1021/jp306160c..

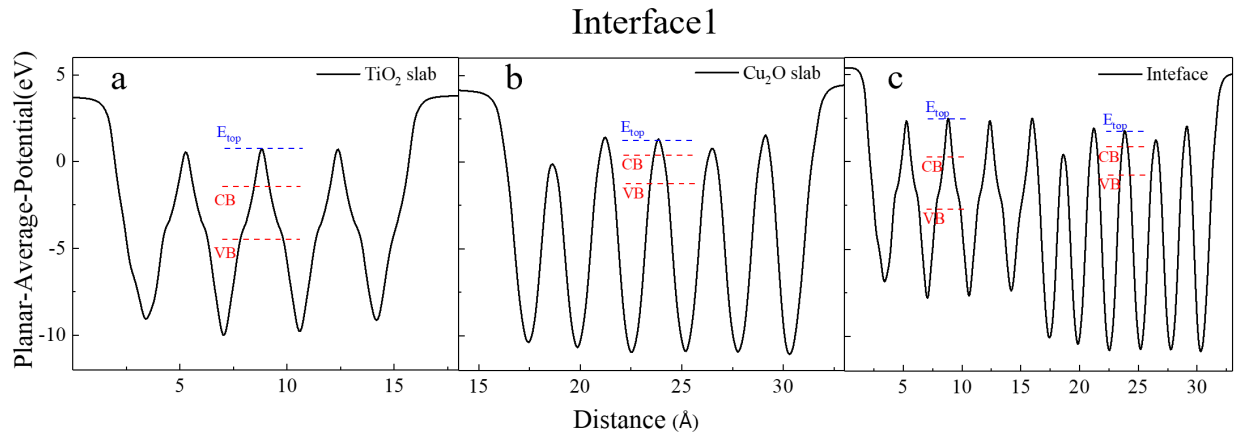

*Figure 13. planar-averaged electrostatic potentials in the x-y plane of (a,b) the isolated slabs with the interface geometries, and (c) of interface 1 as a function of the distance along the z-direction, normal to the interface.*

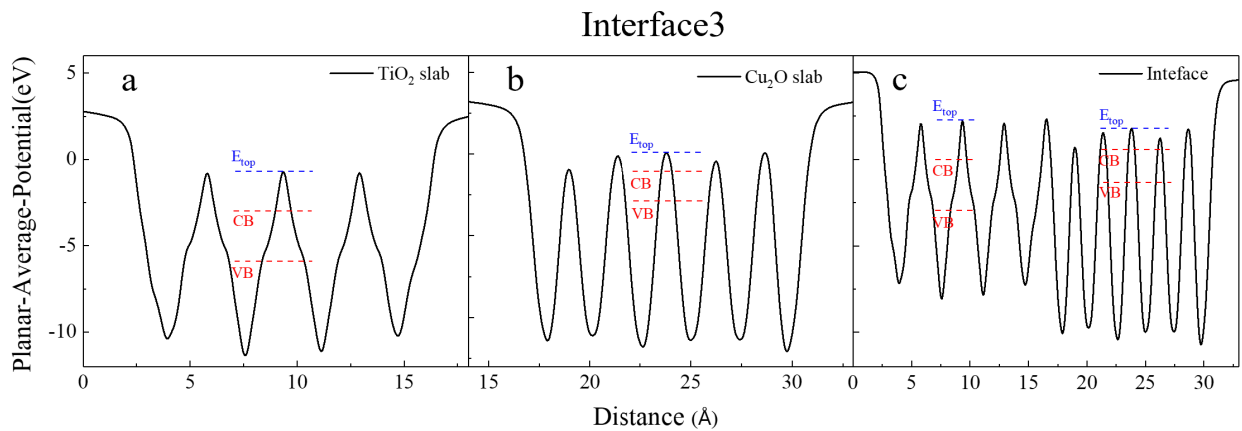

*Figure 14. planar-averaged electrostatic potentials in the x-y plane of (a,b) the isolated slabs with the interface geometries, and (c) of interface 3 as a function of the distance along the z-direction, normal to the interface.*

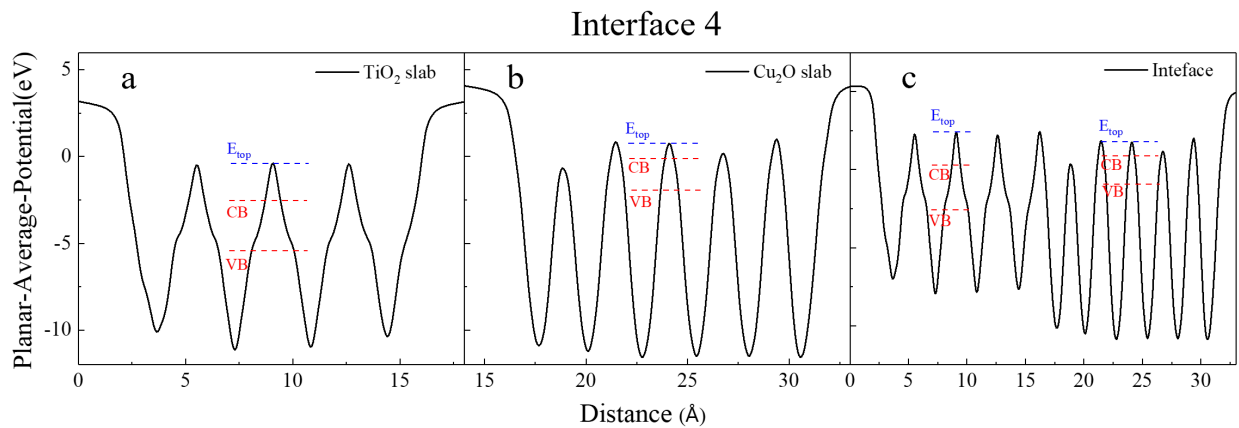

*Figure 15. planar-averaged electrostatic potentials in the x-y plane of (a,b) the isolated slabs with the interface geometries, and (c) of interface 4 as a function of the distance along the z-direction, normal to the interface.*

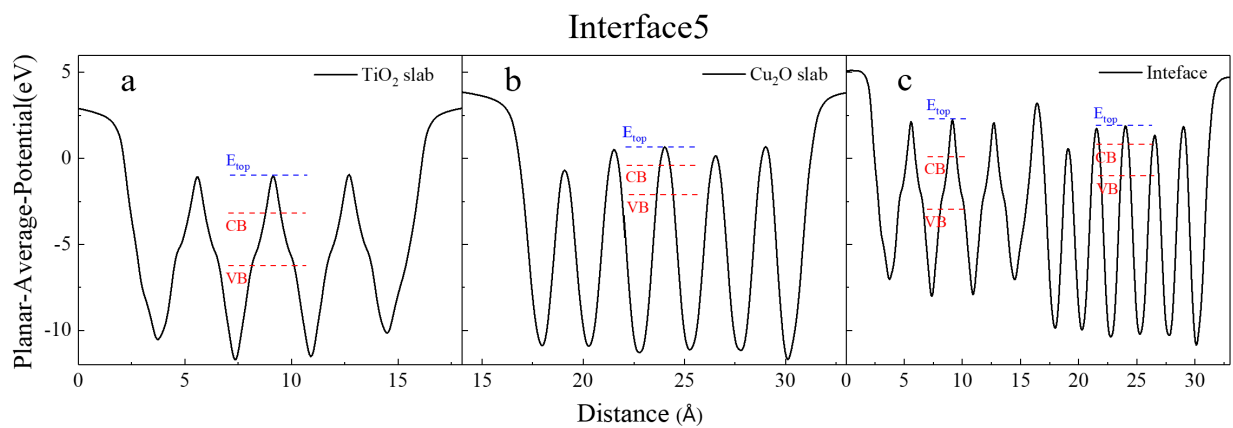

*Figure 16. planar-averaged electrostatic potentials in the x-y plane of (a,b) the isolated slabs with the interface geometries, and (c) of interface 5 as a function of the distance along the z-direction, normal to the interface.*

## Interface6

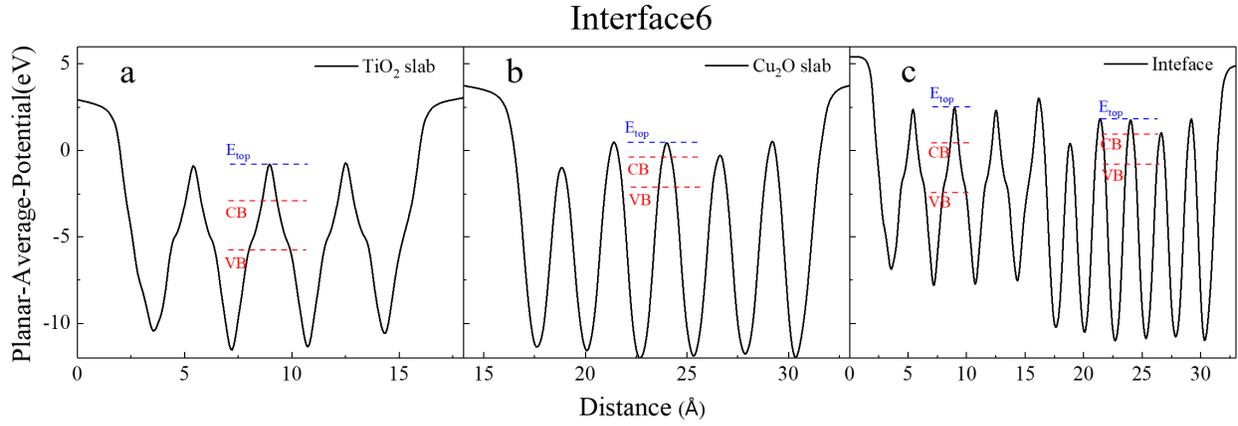

*Figure 17. planar-averaged electrostatic potentials in the x-y plane of (a,b) the isolated slabs with the interface geometries, and (c) of interface 6 as a function of the distance along the z-direction, normal to the interface.*

## Interface7

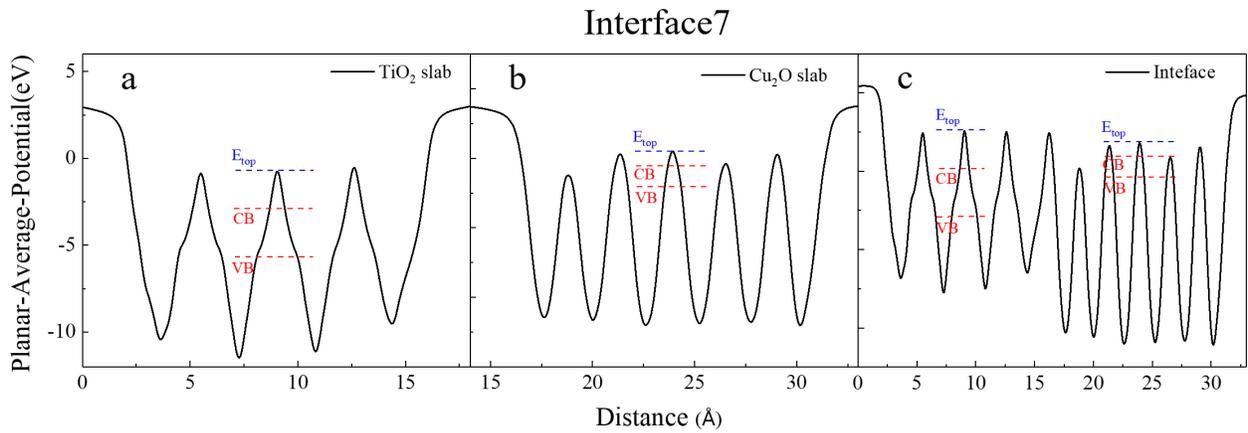

*Figure 18. planar-averaged electrostatic potentials in the x-y plane of (a,b) the isolated slabs with the interface geometries, and (c) of interface 7 as a function of the distance along the z-direction, normal to the interface.*

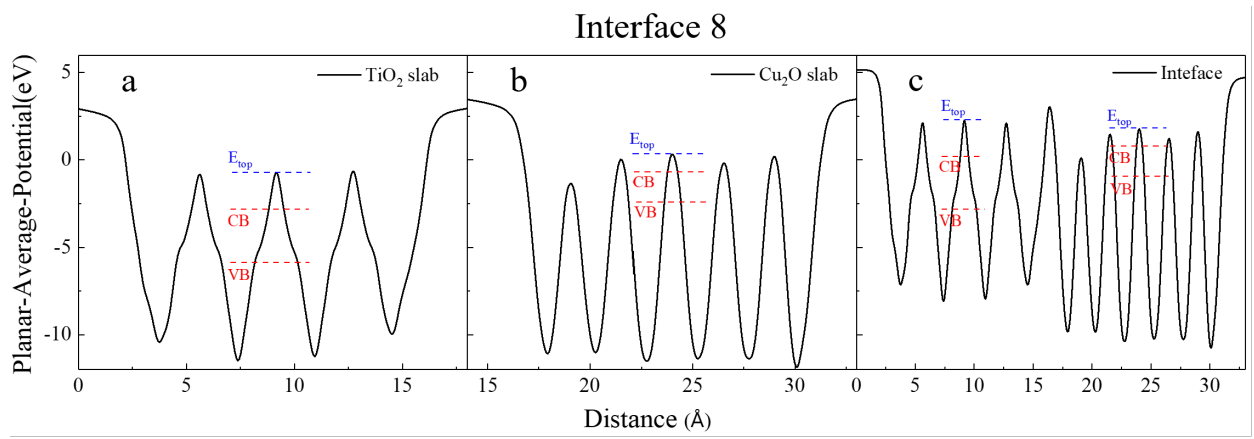

*Figure 19. planar-averaged electrostatic potentials in the x-y plane of (a,b) the isolated slabs with the interface geometries, and (c) of interface 8 as a function of the distance along the z-direction, normal to the interface.*

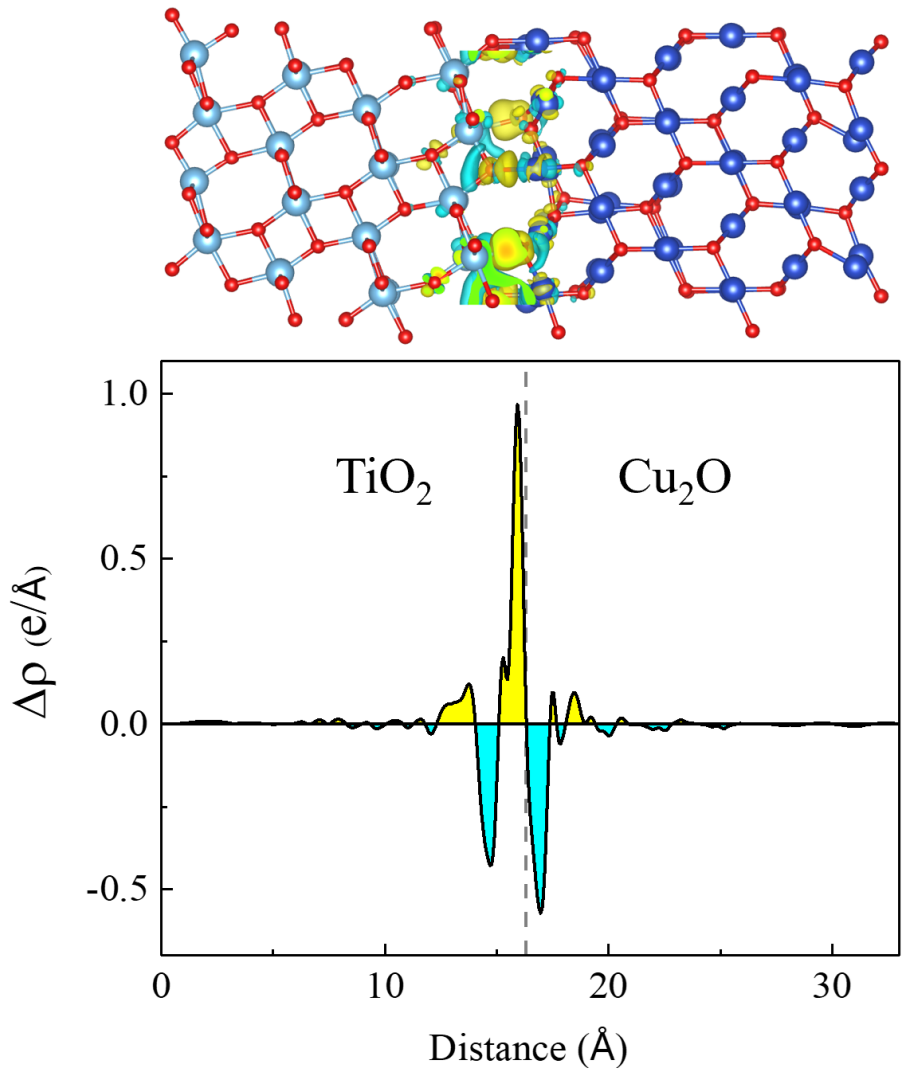

*Figure 20. Charge density difference of interface 1. Yellow: charge accumulation. Cyan: charge depletion. The dashed line represents the midpoint layer of the interface.*

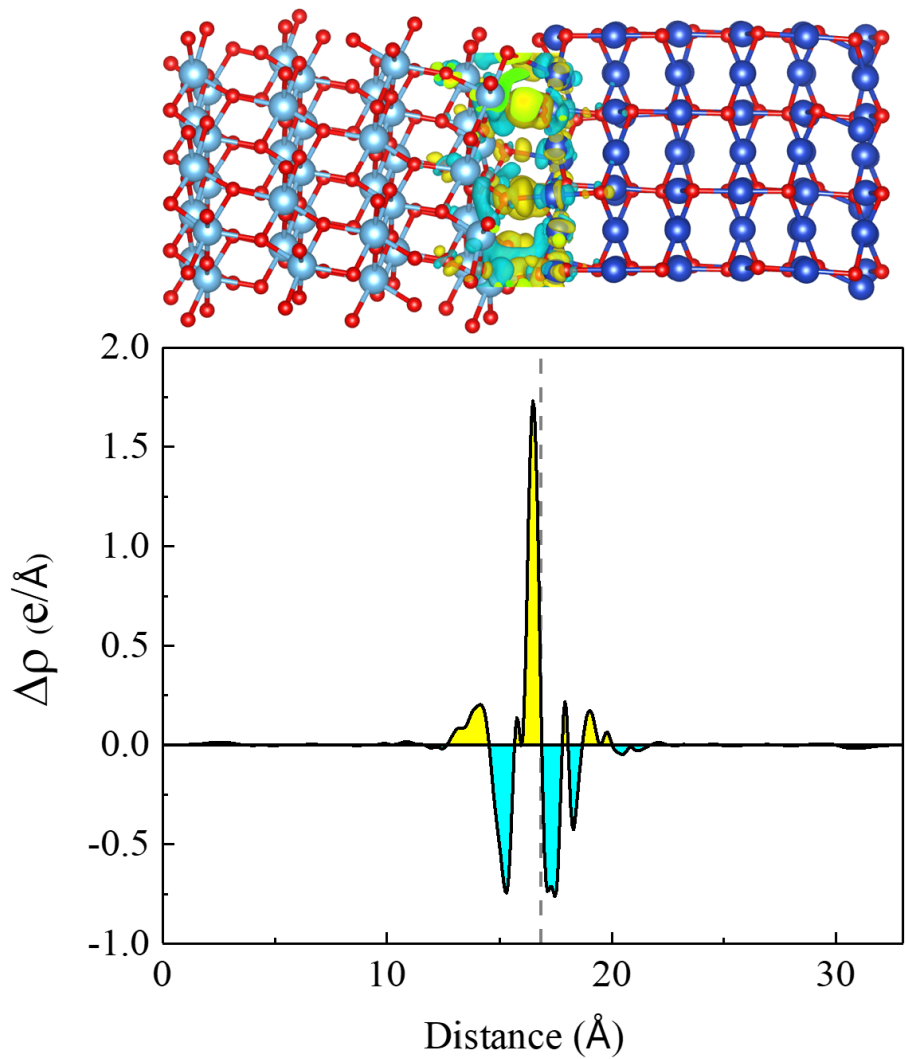

*Figure 21. Charge density difference of interface 3. Yellow: charge accumulation. Cyan: charge depletion. The dashed line represents the midpoint layer of the interface.*

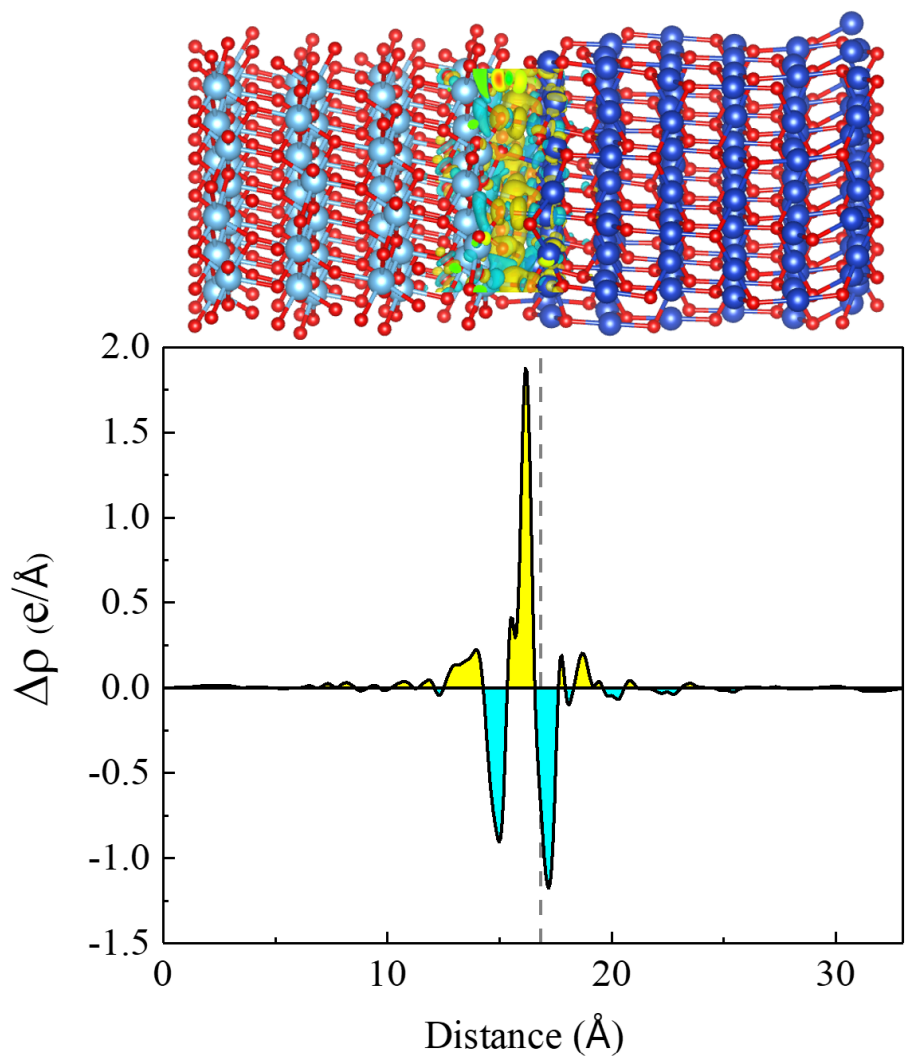

*Figure 22. Charge density difference of interface 4. Yellow: charge accumulation. Cyan: charge depletion. The dashed line represents the midpoint layer of the interface.*

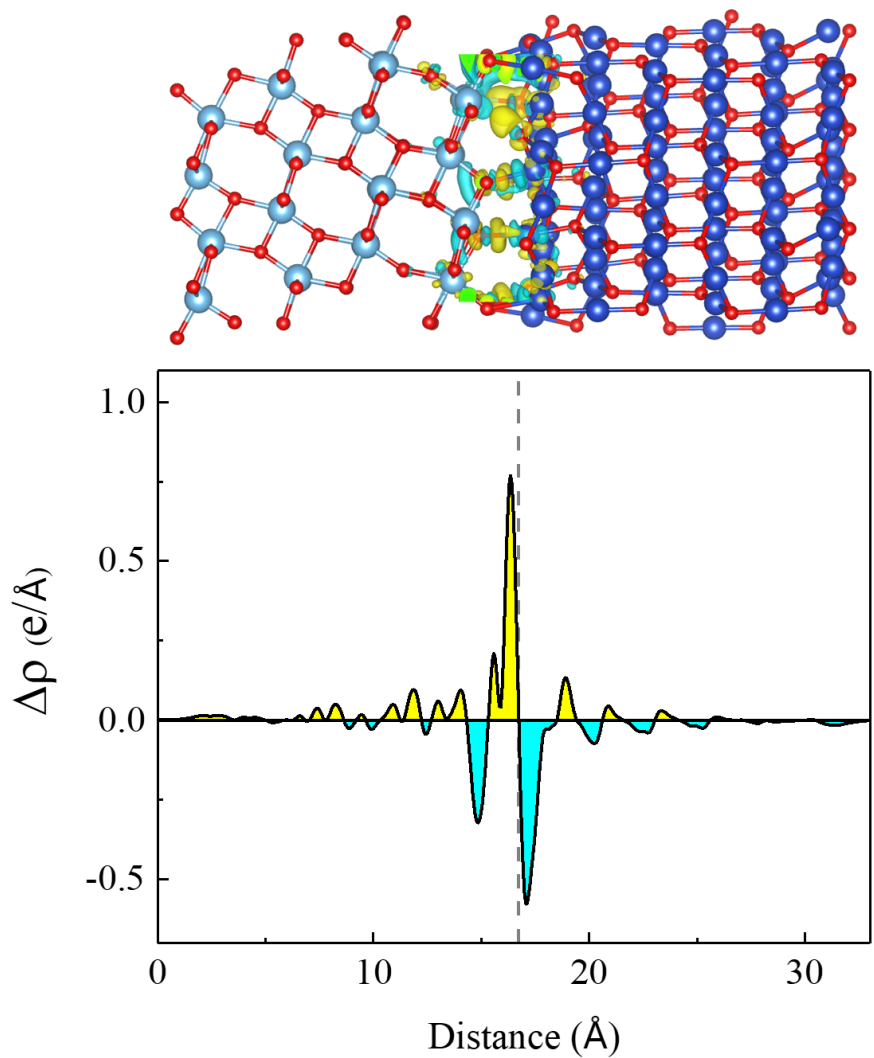

*Figure 23. Charge density difference of interface 5. Yellow: charge accumulation. Cyan: charge depletion. The dashed line represents the midpoint layer of the interface.*

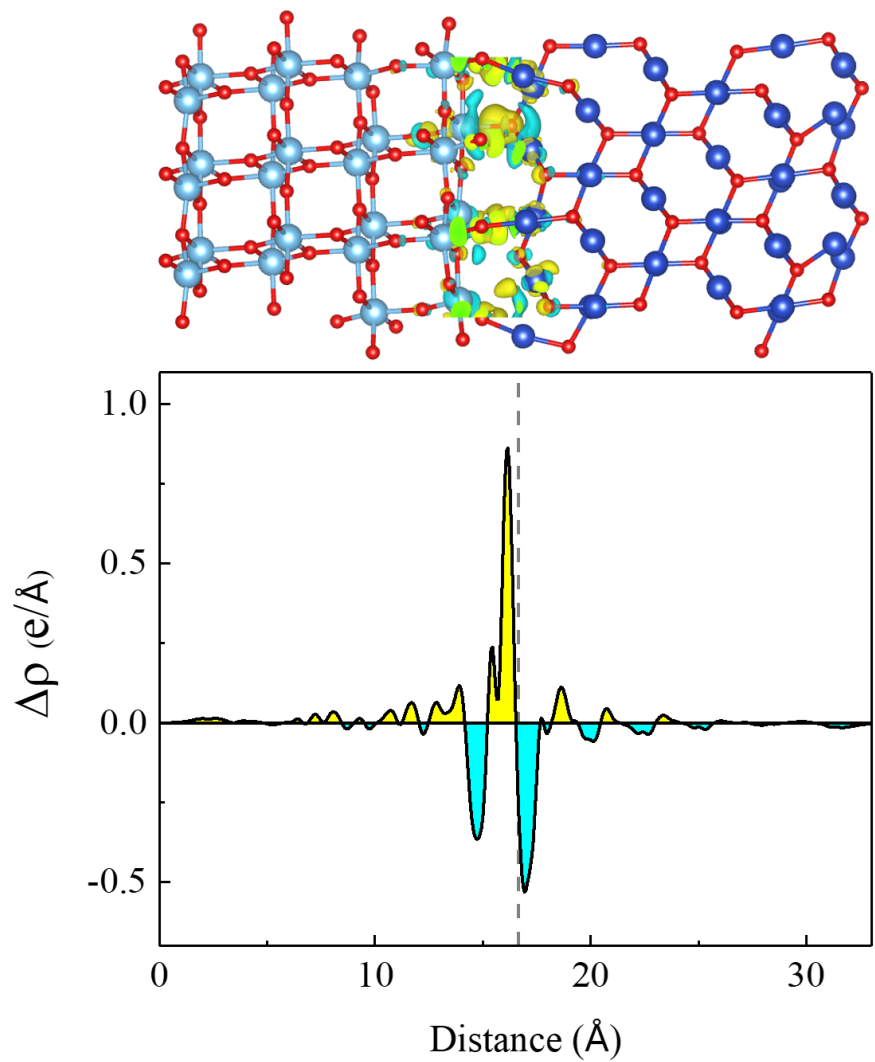

*Figure 24. Charge density difference of interface 6. Yellow: charge accumulation. Cyan: charge depletion. The dashed line represents the midpoint layer of the interface.*

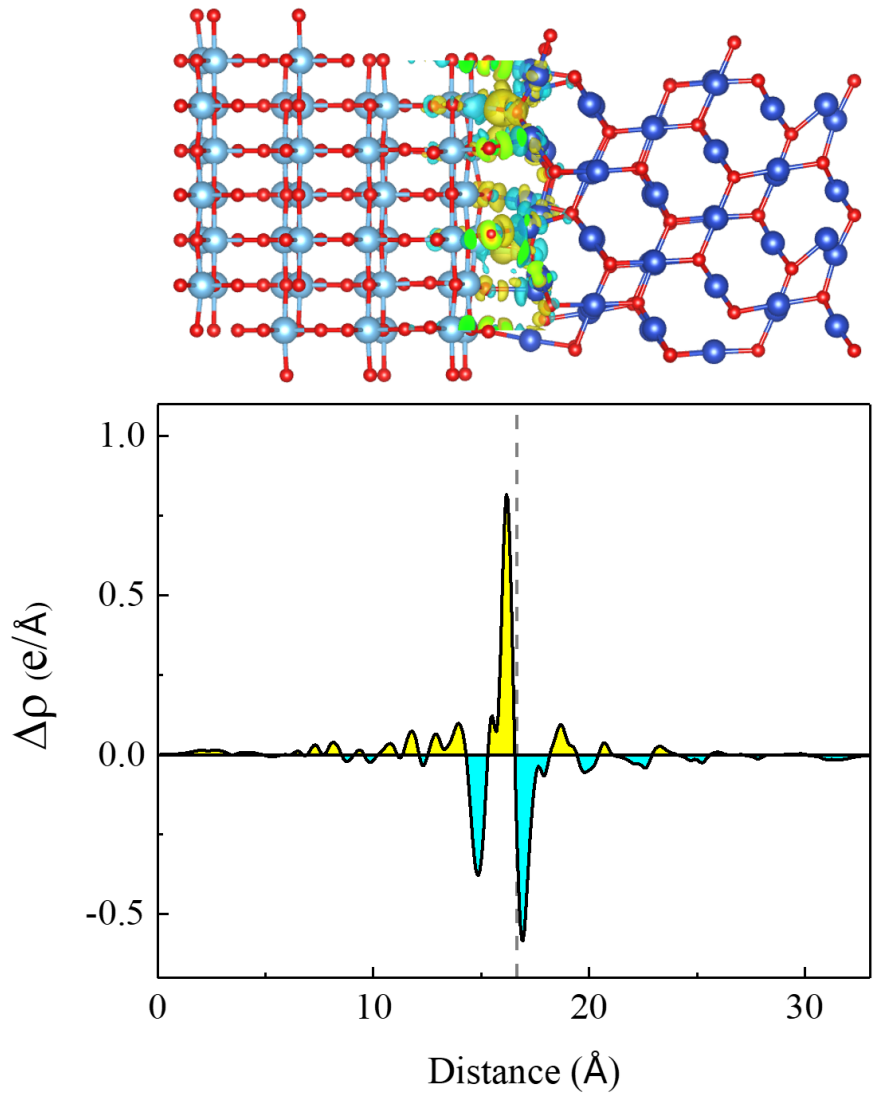

*Figure 25. Charge density difference of interface 7. Yellow: charge accumulation. Cyan: charge depletion. The dashed line represents the midpoint layer of the interface.*

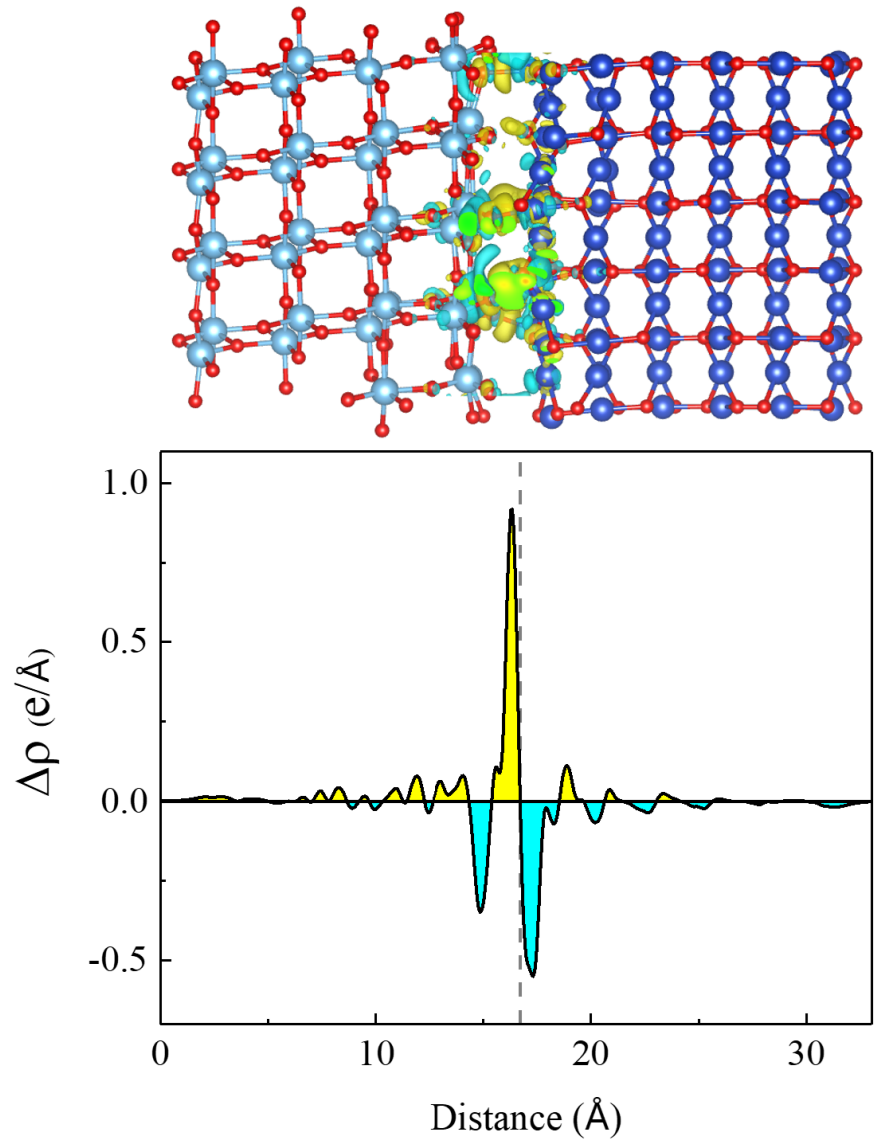

*Figure 26. Charge density difference of interface 8. Yellow: charge accumulation. Cyan: charge depletion. The dashed line represents the midpoint layer of the interface.*

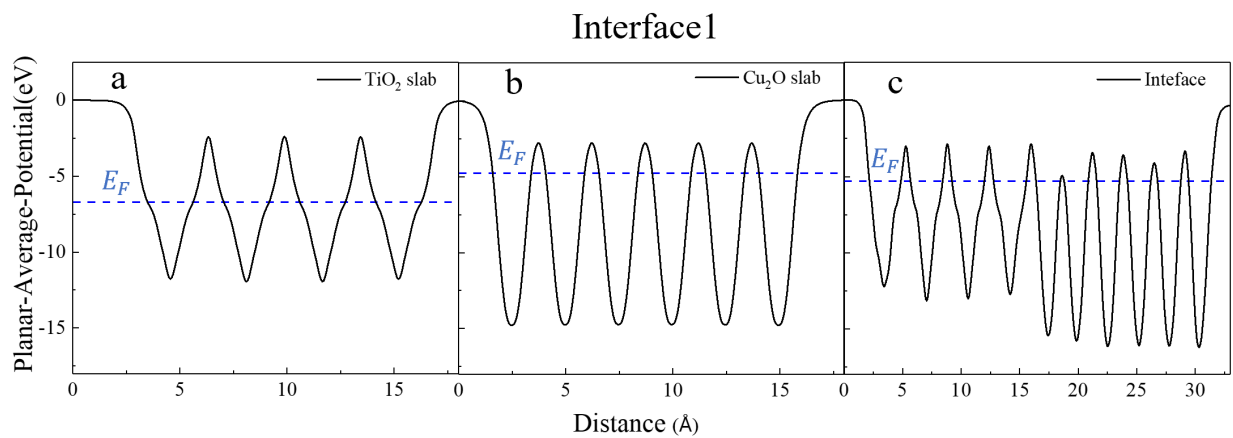

*Figure 27. Fermi energy assessment of interface 1 (a,b) before and (c) after contact*

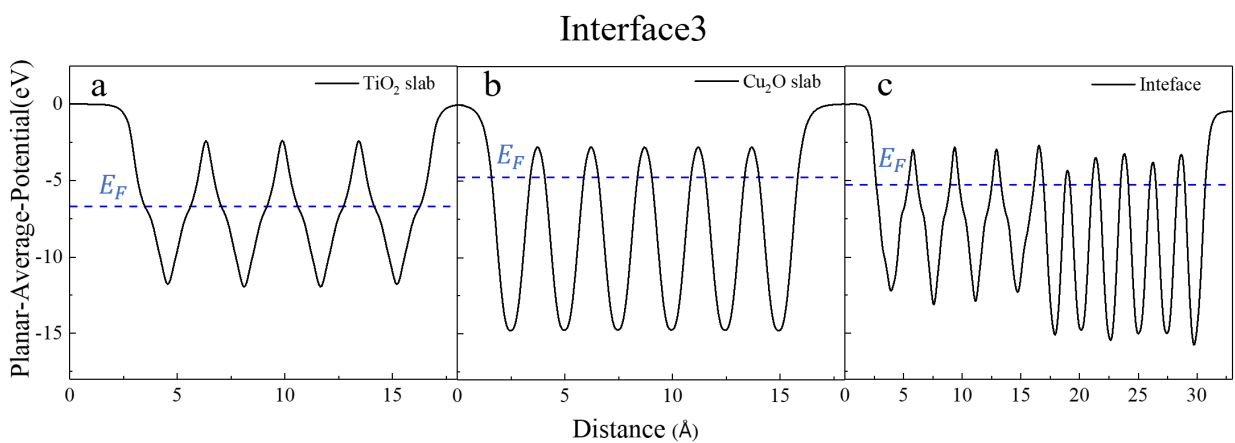

*Figure 28. Fermi energy assessment of interface 3 (a,b) before and (c) after contact*

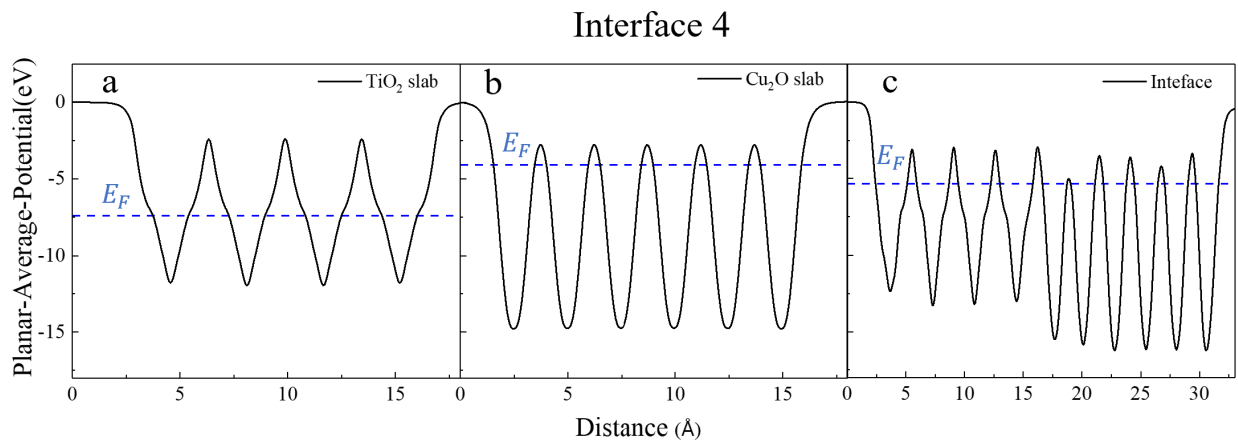

*Figure 29. Fermi energy assessment of interface 4 (a,b) before and (c) after contact*

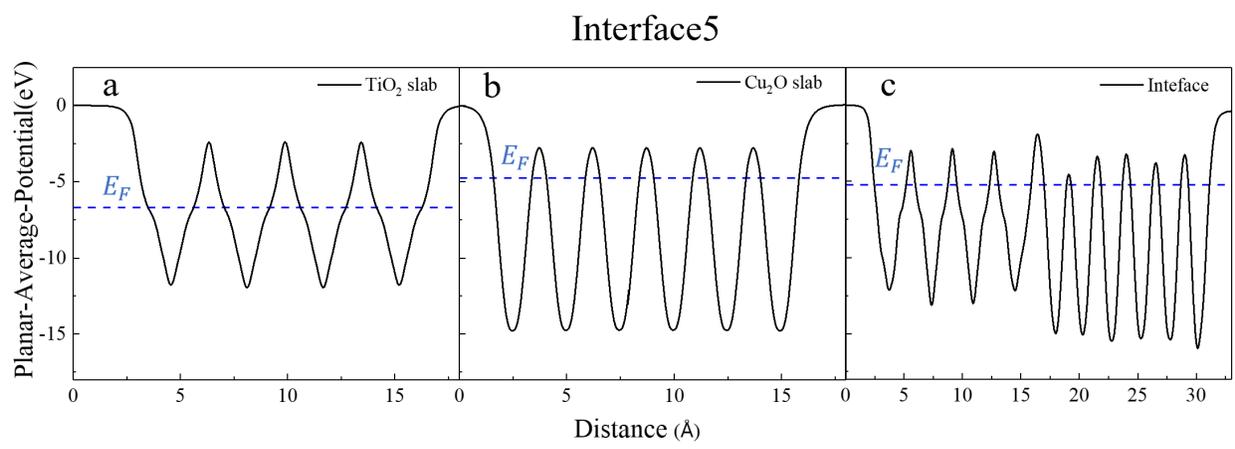

*Figure 30. Fermi energy assessment of interface 5 (a,b) before and (c) after contact*

## Interface6

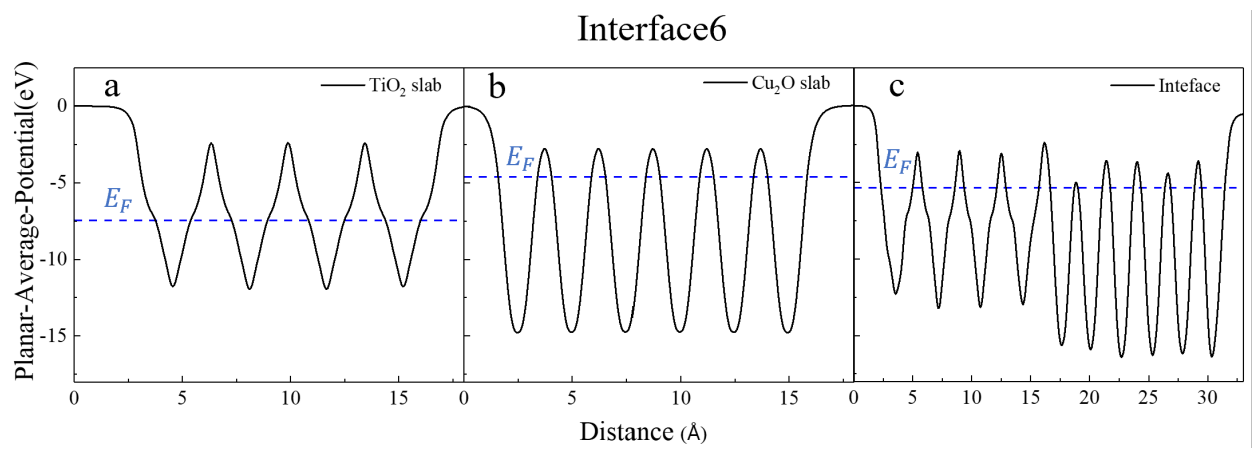

*Figure 31. Fermi energy assessment of interface 6 (a,b) before and (c) after contact*

## Interface7

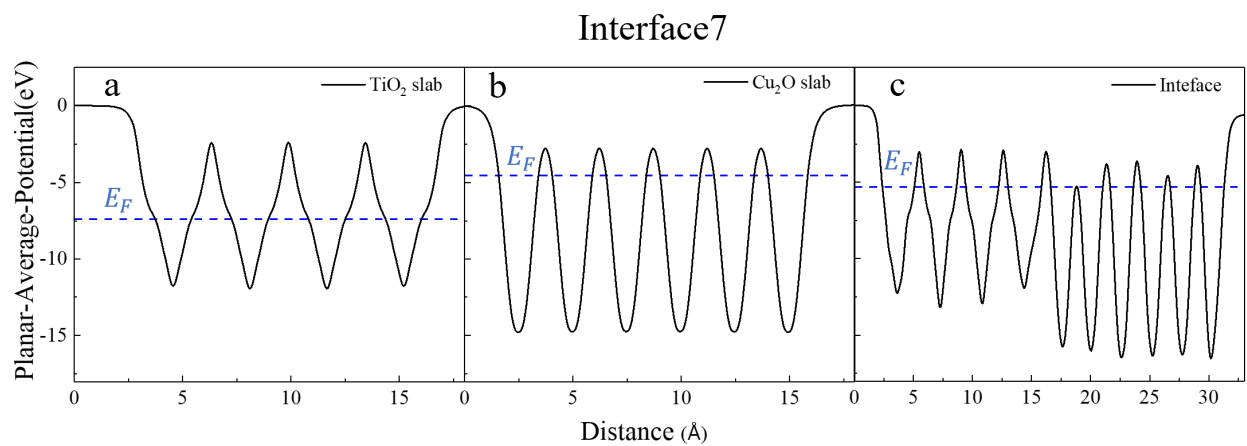

*Figure 32. Fermi energy assessment of interface 7 (a,b) before and (c) after contact*

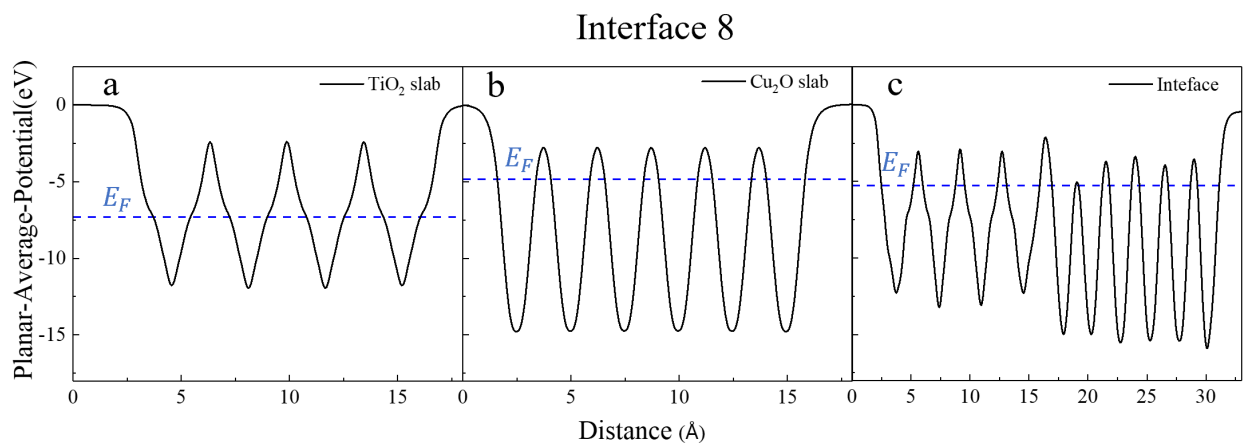

*Figure 33. Fermi energy assessment of interface 8 (a,b) before and (c) after contact*